\documentclass[aps,prd,showpacs]{revtex4}

\usepackage{amsmath,amssymb,amsfonts,amsthm,bbm} 
\usepackage{color}
\usepackage{graphicx}
\usepackage{pstricks,pstricks-add}
\usepackage{pst-plot}
\usepackage[hang,nooneline]{subfigure}
\usepackage{slashbox}

\definecolor{dark-green}{rgb}{0,0.7,0}
\definecolor{dark-blue}{rgb}{0,0.2,0.5}
\definecolor{med-blue}{rgb}{0,0.7,1}
\definecolor{mblue}{rgb}{0,0.2,1}
\definecolor{cnc}{rgb}{0.8,0,0}
\definecolor{light-red}{rgb}{1,0.8,0.8}
\definecolor{dark-yellow}{rgb}{1,0.8,0}
\definecolor{light-blue}{rgb}{0.8,0.9,1}
\definecolor{grey}{rgb}{0.211,0.211,0.211}
\definecolor{verylight-blue}{rgb}{0.93,0.95,1}
\definecolor{light-yellow}{rgb}{1,0.9,0.8}

\begin{document}

\title{The complete set of solutions of the geodesic equations in the space-time of a Schwarzschild black hole pierced by a cosmic string}

\author{Eva Hackmann $^{(a)}$}
\email{hackmann@zarm.uni-bremen.de}
\author{Betti Hartmann $^{(b)}$ }
\email{b.hartmann@jacobs-university.de}
\author{Claus L{\"a}mmerzahl $^{(a)}$}
\email{laemmerzahl@zarm.uni-bremen.de}
\author{Parinya Sirimachan $^{(b)}$}
\email{p.sirimachan@jacobs-university.de}

\affiliation{
$(a)$ ZARM, Universit\"at Bremen, Am Fallturm, 28359 Bremen, Germany\\
$(b)$ School of Engineering and Science, Jacobs University Bremen, 28759 Bremen, Germany}
\date\today

\begin{abstract}

We study the geodesic equations in the space-time of a Schwarzschild black hole pierced by an infinitely thin cosmic string and give the complete set of analytical solutions of these equations for massive and massless
particles, respectively.  The solutions of the geodesic equations can be classified according to the particle's energy and angular momentum, the ratio between the component of the angular momentum aligned with the axis of the string
and the total angular momentum, the deficit angle of the space-time and as well the horizon radius (or mass) of the
black hole. For bound orbits of massive test particles we calculate the perihelion shift, we discuss light deflection and comment on the Newtonian limit.

\end{abstract}


\maketitle

\section{Introduction}
The motion of test particles (both massive and massless) provides the only experimentally feasible way to study the gravitational fields of objects such as black holes.
Predictions about observable effects (light deflection, gravitational
time--delay, the perihelion shift and the Lense-Thirring effect) can be
made and compared with observations. Geodesics in black hole space-times in 4 
dimensional Schwarzschild space--time \cite{hagihara} and Kerr
and Kerr--Newman space--time \cite{chandra}
have been discussed extensively.
This has been extended to the cases of Schwarzschild--de--Sitter space-times \cite{hl1} as well as to spherically symmetric higher dimensional
space--times \cite{hackmann08}. Recently also the general solution
to the geodesic equation in 4 dimensional Kerr--de--Sitter \cite{hackmannetal1} and even general Plebanski--Demianski space--times
without acceleration has been found \cite{hackmannetal2}.

Cosmic strings have gained a lot of renewed interest over the past years
due to their possible connection to string theory \cite{polchinski}.
These are topological defects \cite{vs} that could have formed in one of the numerous phase transitions in the early universe due to the Kibble mechanism.
Inflationary models resulting from string theory (e.g. brane inflation)
predict the formation of cosmic string networks at the end of inflation \cite{braneinflation}.

Different space-times containing cosmic strings have been discussed in the past.
This study has mainly been motivated by the pioneering work of Bach and Weyl \cite{bw} 
describing a pair of black holes held apart by an infinitely thin strut.
This solution has later been reinterpreted in terms of cosmic strings describing
a pair of black holes held apart by two cosmic strings extending to infinity in opposite direction.
Consequently, a cosmic string piercing a Schwarzschild black hole has also been discussed, both in the thin string limit \cite{afv} -- where an analytic solution can be given --
as well as using the full U(1) Abelian-Higgs model \cite{dgt,agk}, where only numerical solutions
are available. In the latter case, these solutions have
been interpreted to represent black hole solutions with long range ``hair'' and are thus counterexamples
to the No hair conjecture which states that black holes are uniquely characterized by
their mass, charge and angular momentum.
Interestingly, the solution found in \cite{afv} is a Schwarzschild solution
which however differs from the standard spherically symmetric case by the replacement of the angular
variable $\phi$  by $\beta\phi$, where the parameter $\beta$ is related to the deficit
angle by $\Delta=2\pi(1-\beta)$. In this sense, the space-time is thus not uniquely determined by the mass, but
is described by the mass {\it and} deficit angle parameter $\beta$.

Schwarzschild black holes pierced by cosmic strings could have 
formed in phase transitions in the early universe. One possibility
would be that the Coulomb field of a charged static, spherically symmetric black hole becomes confined within a flux tube during such a phase transition. 
Interestingly, this space--time has also been used to describe the exterior  space--time of the sun taking into account departures from
perfect spherical symmetry \cite{fbl}.

In order to understand details of gravitational fields of 
massive objects and to be able to predict observational consequences, it is
important to understand how test particles move in these space-times.

Geodesics in the space--time of a Schwarzschild black hole pierced by
a straight cosmic string have first been investigated in \cite{ag}.
Spherical geodesics, i.e. geodesics in 3 spatial dimensions with constant radius $r$ as well as perturbations about spherical orbits in this space--time have been discussed in \cite{gm} and it has been observed that the
angular momentum precesses around the symmetry axis of the cosmic string. The geodesics have
also been discussed in \cite{cb} using the Hamilton-Jacobi formalism.
However, neither \cite{gm} nor \cite{cb} provide a systematic study of all possible geodesics. The crucial point about the solutions of the geodesic
equations is that elliptic integrals have to be solved which
was neither attempted in \cite{gm} nor \cite{cb}.

The aim of this paper is to determine the {\it complete set} of
analytic solutions of the geodesic equations in the space--time of
a Schwarzschild black hole pierced by a cosmic string and to derive analytical expressions for observable effects which can be used for astrophysical searches
for such cosmic strings.

Our paper is organized as follows: in Section II, we give the geodesic equations and discuss the effective
potential.
In Section III, we classify the solutions and give examples for each class. In particularly we
discuss the effect of the conical deficit on the geodesics. Section IV contains a short discussion about the Newtonian limit, while we conclude in Section V.

\section{The geodesic equations}
We consider the geodesic equation
\begin{equation}
 \frac{d^2 x^{\mu}}{ds^2} + \Gamma^{\mu}_{\rho\sigma} \frac{dx^{\rho}}{ds}\frac{dx^{\sigma}}{ds}=0  \ ,
\end{equation}
where $\Gamma^{\mu}_{\rho\sigma}$ denotes the Christoffel symbol given by
\begin{equation}
 \Gamma^{\mu}_{\rho\sigma}=\frac{1}{2}g^{\mu\nu}\left(\partial_{\rho} g_{\sigma\nu}+\partial_{\sigma} g_{\rho\nu}-\partial_{\nu} g_{\rho\sigma}\right)
\end{equation}
and $s$ is an affine parameter such that for time--like geodesics $ds^2=g_{\mu\nu}dx^{\mu} dx^{\nu}$ corresponds to proper time.
The explicit form of the metric that we are studying in this paper is the 
metric of a Schwarzschild black hole pierced by a cosmic string \cite{afv}~:
\begin{eqnarray}
\label{metric}
ds^{2}=\Sigma dt^{2}-\Sigma^{-1}dr^{2}-r^{2}(d\theta^{2}+\beta^{2}\sin^{2}\theta d\phi^{2}) \  \ \ \ \ {\rm with} \ \ \ \ \Sigma=1-\frac{2m}{r}   \ .
\end{eqnarray}
This metric describes a spherically symmetric static space-time with a 
conical deficit angle given by $2\pi(1-\beta)$ and an event horizon
at the Schwarzschild radius $r_s=2m$. The deficit angle $\Delta$ is directly proportional
to the energy per unit length $\mu$, which itself is equal to the tension of the string:
$\Delta=2\pi(1-\beta)=8\pi G \mu$.   
In addition we have $m=MG$, where $G$ is Newton's constant. Note that $M$ is a parameter that is related
to the physical mass $M_{\rm phys}$ of the black hole by $M_{\rm phys}=\beta M$ \cite{gm} and can hence
only be interpreted as the mass  of the black hole if $\beta=1$. $M_{\rm phys}$ is observable through
the measurement of the event horizon radius by $M_{\rm phys}= (\beta r_s)/(2G)$.

The Lagrangian $\mathcal{L}$ for a point particle in the space--time (\ref{metric}) reads~:
\begin{eqnarray}
\label{lagrangian}
\mathcal{L}&=&\frac{1}{2}g_{\mu\nu}\frac{dx^{\mu}}{ds}\frac{dx^{\nu}}{ds}=\frac{1}{2}\varepsilon\\
&=&\frac{1}{2}\left[\left(1-\frac{2m}{r}\right)\left(\frac{dt}{ds}\right)^{2}-
\left(1-\frac{2m}{r}\right)^{-1}\left(\frac{dr}{ds}\right)^{2}-r^{2}\left(\left(\frac{d\theta}{ds}\right)^{2}+\beta^{2}\sin^{2}\theta\left(\frac{d\phi}{ds}\right)^{2}\right)\right]\nonumber  \ ,  \end{eqnarray}
where $\varepsilon=0$ for massless particles and $\varepsilon=1$ for massive particles, respectively. 

The constants of motion are the energy $E$ and the component of the angular
momentum of the particle that is aligned with the axis of the string (here the $z$-axis) $L_z$ \cite{gm}~:
\begin{eqnarray}
E&=&\left(1-\frac{2m}{r}\right)\frac{dt}{ds}\quad={\rm constant}\label{dott} \ \ , \ \\
L_z&=&r^{2}\beta\sin^{2}\theta\frac{d\phi}{ds}\quad={\rm constant}\label{dotphi}  \ .
\end{eqnarray}
In addition
\begin{eqnarray}\label{totalL}
\vert \vec{L}\vert^2\equiv L^{2}&=&\left(\frac{d\theta}{ds}\right)^{2}r^{4}+\frac{L_{z}^{2}}{\sin^{2}\theta}={\rm constant} \ .
\end{eqnarray}
Here $\vec{L}=(L_x,L_y,L_z)$ is the angular momentum vector, where 
\begin{equation}
 L_x=- r^2 \sin(\beta\phi) \frac{d\theta}{ds} - \beta r^2 \cos(\beta\phi) \cos\theta\sin\theta \frac{d\phi}{ds}
\end{equation}
and
\begin{equation}
 L_y=r^2 \cos(\beta\phi)\frac{d\theta}{ds} - \beta r^2 \sin(\beta\phi)\cos\theta\sin\theta\frac{d\phi}{ds}  \ .
\end{equation}
The modulus of the angular momentum
is always conserved, however the direction of the angular momentum is only conserved for $\beta=1$, i.e.
in the ``pure'' Schwarzschild case.
 
Note that if we had used the Hamilton-Jacobi formalism here, the integration
constant appearing in the separation of the equations --the so-called Carter constant-- would
have been equal to $L^2-L_z^2$.

From variation of (\ref{lagrangian}) and using the constants of motion, we obtain the geodesic equations~:  
\begin{eqnarray}\label{tcompo}
\dot{t}^2&=&E^{2}\left(1-\frac{2m}{r}\right)^{-2} \ , \\
\label{rcompo}
\dot{r}^2&=&E^{2}-\left(\frac{L^{2}}{r^{2}}+\varepsilon\right)\left(1-\frac{2m}{r}\right) 
 \ , \\
\label{thetacompo}
\dot{\theta}^{2}&=&\frac{L^{2}}{r^{4}}-\frac{L_{z}^{2}}{r^{4}\sin^{2}\theta}   \ ,  \\
\label{phicompo}
\dot{\phi}^{2}&=&\frac{L^{2}_{z}}{\beta^{2} r^{4}\sin^{4}\theta}  \ ,
\end{eqnarray}
where here and in the following the dot denotes the derivative with respect to the affine parameter $s$.
\subsection{Effective potentials}
Equations (\ref{rcompo}) and (\ref{thetacompo}) can be rewritten in terms of
effective potentials as follows:
\begin{equation}
\label{effective}
 \frac{1}{2}\dot{r}^2+V_{\rm eff}(r)= \frac{E^2-\varepsilon}{2} \ \ \ , \ \ \
r^4\dot{\theta}^2+V_{\rm eff}(\theta)= L^2  \ ,
\end{equation}
where
\begin{equation}
 V_{\rm eff}(r)=-\varepsilon\frac{m}{r}+\frac{L^2}{2r^2} - \frac{L^2 m}{r^3} \ \ , \ \
V_{\rm eff}(\theta)=\frac{L_z^2}{\sin^2\theta}  \ .
\end{equation}
The effective potential $V_{\rm eff}(r)$ is exactly the same as that in the ``pure''
Schwarzschild case. Since $L^2 \geq V_{\rm eff}(\theta)$, the $\theta$ motion is
restricted:
\begin{equation}
 \arcsin\left(\frac{L_z}{L}\right) \le \theta \le \pi-\arcsin\left(\frac{L_z}{L}\right)  \ .
\end{equation}
Apparently, for $L_z=L$, the motion occurs in the equatorial plane, i.e. $\theta=\pi/2$.

\subsection{$r(\theta)$ and $r(\phi)$ motion}


In the following, we will be mainly interested in the radial motion, i.e.
we will solve the geodesic equations for $r(\theta)$ and $r(\phi)$.
In order to do that we first have to eliminate the angular
variable $\theta$ from (\ref{phicompo}).
By dividing (\ref{thetacompo}) by ({\ref{phicompo}}) we find \cite{gm}:
\begin{eqnarray}
\frac{d\theta}{d\phi}&=&\beta\sin\theta\sqrt{k^{2}\sin^{2}\theta-1}  \ ,
\end{eqnarray}  
where $k^{2}$=$(L/L_{z})^{2}$. This can be solved to give \cite{gm}:
\begin{equation}
\label{thetaphi}
 \cot^2\theta=(k^2-1)\sin^2(\beta\phi)   \ .
\end{equation}
Using (\ref{thetaphi}), (\ref{rcompo}), (\ref{phicompo}) and (\ref{thetacompo}) we find:
\begin{equation}
\label{rtheta}
\left(\frac{dr}{d\theta}\right)^2=\frac{E^2-\left(\frac{L^2}{r^2}+\varepsilon\right)\Sigma}{L^2\sin^2\theta-L_z^2} r^4 \sin^2\theta
\end{equation}
and
\begin{equation}
\label{rphi}
\left(\frac{dr}{d\phi}\right)^2=\frac{E^2-\left(\frac{L^2}{r^2}+\varepsilon\right)\Sigma}{L_z^2}
\frac{\beta^2 r^4}{\left((k^2-1)\sin^2(\beta\phi)+1\right)^2}   \ .
\end{equation}
Orbits with $\phi={\rm constant}$ are exactly the same as in the Schwarzschild case $(\beta=1$), 
while for orbits with $\phi\neq {\rm constant}$ the presence of the deficit angle influences the shape of the orbits
significantly. The orbits are in general non-planar (except for the case $L_z=L$) and lie in a plane
that has $\vec{L}$ as its normal. Since $\vec{L}$ is not conserved, this plane precesses \cite{gm}.
 
In \cite{gm}, the simplest case of non-planar orbits, namely spherical orbits and perturbations around spherical orbits have been discussed.
Here, we want to give the complete set of possible orbits in this space--time and discuss
all related observables.

\subsection{Classification of solutions}
\begin{figure}[htbp]
\begin{center}
\resizebox{!}{2in}{\includegraphics{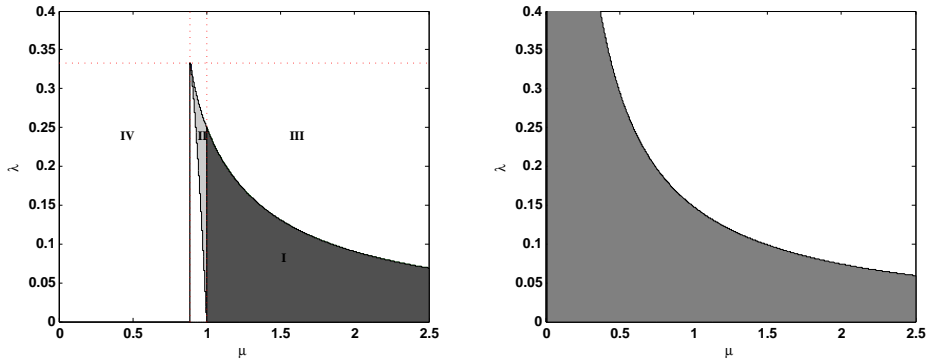}}
\end{center}
\caption{The regions corresponding to positive, negative and zero discriminant $D$ in the $\lambda-\mu$ plane for
massive test particles (left) and massless test particles (right), respectively. For massive test particles
the shaded regions I and II correspond to positive discriminant ($D >0$), the unshaded regions III and IV
correspond to negative discriminant ($D <0$) and the boundary between the two regions to zero discriminant $D=0$.
The vertical line at $\mu=1$ separates the motion of massive particles ($\varepsilon=1$) with $\mu-\varepsilon >0$ and
$\mu-\varepsilon < 0$, respectively. For massless test particles, the shaded region corresponds to $D>0$, the unshaded
region to $D <0$ and the full boundary of the $D>0$ region to $D=0$.
}\label{ldmuplot}
\end{figure}

Using the new variable $z=\frac{2m}{r}-\frac{1}{3}$ and introducing
\begin{equation}
 \lambda=\frac{4m^2}{L^2} \ \ , \ \ \mu=E^2 \ \ , \ \ k=\frac{L}{L_z}
\end{equation}
we find from (\ref{rtheta}) and (\ref{rphi}) using (\ref{thetaphi}):
\begin{eqnarray}
\label{differential1}
\frac{dz}{\sqrt{P(z)}} =\frac{1}{2} \left(1-\frac{1}{k^2\sin^2\theta}\right)^{-1/2} d\theta  \ ,
\end{eqnarray}
\begin{eqnarray}
\label{differential2} 
\frac{dz}{\sqrt{P(z)}}  =  \frac{1}{2}\beta k \frac{1}{(k^2-1)\sin^2(\beta\phi)+1} d\phi  \ , 
\end{eqnarray}
where the third order polynomial $P(z)$ is given by
\begin{equation}
P(z)= 4z^3 - g_2 z - g_3 \ \ , \ \ g_2=4\left(\frac{1}{3}-\varepsilon\lambda\right) \ \ , \ \ g_3= 4\left(\frac{2}{27} + \frac{2}{3} \varepsilon\lambda -\lambda \mu\right)
\end{equation}
This is exactly the polynomial that appears in the case of the Schwarzschild solution and the classification of solutions is analogue. The main difference
to the ``pure'' Schwarzschild case is that we have non-trivial
integrals on the rhs of (\ref{differential1}), (\ref{differential2}). For a given $k$ and $\beta$, the solutions can hence be classified according to the choice of
$\mu$ and $\lambda$. Obviously, there are only solutions for $P(z) > 0$. Hence, we can classify 
solutions according to the zeros of the characteristic polynomial $P(z)$ and integrate (\ref{differential1}), (\ref{differential2}) for those $z$ for which $P(z) > 0$. Note that in addition we can only integrate
for $z \ge -1/3$ such that $r\ge 0$.
Depending on the sign of the discriminant of $P(z)$ which
is given by $D=g_2^3 - 27 g_3^2$, the polynomial has either three real zeros $(D >0)$, one real zero ($D < 0$)
or up to two real zeros for $D=0$ which we will denote by $e_1$, $e_2$ and $e_3$ where
$e_1 \le  e_2 \le e_3$ in the following.
This is illustrated in Fig.\ref{ldmuplot} (left) for massive test particles ($\varepsilon=1$), where the shaded regions I and II correspond to positive
$D$, the boundary of the shaded region corresponds to $D=0$ and the unshaded regions
III and IV correspond to negative $D$. Note that $D > 0$ only if $\lambda < 1/3$ and $\mu > 8/9$.
We also give the corresponding plot for massless test particles ($\varepsilon=0$) in Fig.\ref{ldmuplot} (right).
The shaded region corresponds to $D>0$, the unshaded region to $D<0$ and the full boundary of the shaded region
to $D=0$. Note that $D >0$ ($D <0$) for $\lambda\mu < 4/27$ ($\lambda\mu > 4/27$) and that
$D=0$ for $\lambda\mu = 4/27$ or $\lambda\mu=0$.

Comparing with (\ref{effective}), $D >0$ corresponds
to the case for which $V_{\rm eff}^{\rm min} < \frac{E^2-\varepsilon}{2} < V_{\rm eff}^{\rm max}$, i.e.
the total ``energy''  $\frac{E^2-\varepsilon}{2}$ of the particle is smaller than the maximum $V_{\rm eff}^{\rm max}$
and larger than the minimum $V_{\rm eff}^{\rm min}$ of the effective potential $V_{\rm eff}(r)$. 
For $D < 0$ then $ \frac{E^2-\varepsilon}{2} < V_{\rm eff}^{\rm min}$ or $\frac{E^2-\varepsilon}{2} > V_{\rm eff}^{\rm max}$, while
for $D=0$, $ \frac{E^2-\varepsilon}{2} = V_{\rm eff}^{\rm min}$ or $\frac{E^2-\varepsilon}{2} = V_{\rm eff}^{\rm max}$.

\section{Solutions of the geodesic equations}
We can integrate (\ref{differential1})
\begin{equation}
\label{rthetarelation}
g(\theta)\equiv\frac{1}{2}\left[\arcsin\left(\frac{\cos\theta}{\sqrt{1-k^{-2}}}\right) - \arcsin\left(\frac{\cos\theta_0}{\sqrt{1-k^{-2}}}\right)\right] = \int_{z_{0}}^{z(\theta)} \frac{dz}{\sqrt{P(z)}}
\end{equation}
and (\ref{differential2})
\begin{equation}
\label{rphirelation}
f(\phi)\equiv -\frac{1}{2}\arctan\left[k\tan(\beta(\phi-\phi_0))\right]= \int_{z_{0}}^{z(\phi)} \frac{dz}{\sqrt{P(z)}}  \ ,
\end{equation}
such that the general solutions for the radial motion in dependence on the angular variables read
\begin{equation}
\label{solutions_r}
 r(\theta)=\frac{2m}{\wp(g(\theta)-{\rm c})+ \frac{1}{3}} \ \ , \ \
r(\phi)=\frac{2m}{\wp(f(\phi)-{\rm c})+ \frac{1}{3}} 
\end{equation}
where $\wp$ is the Weierstrass elliptic function and $g(\theta)$ and $f(\phi)$ are the functions appearing in (\ref{rthetarelation}) and (\ref{rphirelation}), respectively. 
Our choice of $\theta_0$ depends on whether the orbit has finite maximal radius or whether it extends to infinity. For orbits with a
finite maximal radius, we find it convenient to choose $\theta_0=\pi-\arcsin(1/k)$ such that
\begin{equation}
 g(\theta)\equiv g_1(\theta)= \frac{1}{2}\left[\arcsin{\left(\frac{\cos{\theta}}{\sqrt{1-k^{-2}}}\right)}+\frac{\pi}{2}\right]  \ ,
\end{equation}
while for orbits that extend to infinity we choose $\theta_0=\pi/2$ such that
\begin{equation}
 g(\theta)\equiv g_2(\theta)=\frac{1}{2}\arcsin\left(\frac{\cos\theta}{\sqrt{1-k^{-2}}}\right)  \ .
\end{equation}
In addition we choose $\phi_0=0$ for all cases such that from here on
\begin{equation}
f(\phi)=-\frac{1}{2}\arctan\left[k\tan(\beta\phi)\right]  \ .
\end{equation}

The constant $c$ in (\ref{solutions_r}) is given by $c=\int\limits_{z_0}^{\infty} \frac{dz}{\sqrt{4z^3 - g_2 z-g_3}}$.
For $D \leq 0$ we can always choose $c=0$. For $D >0$  we choose
$c=0$ for $z_0=\infty$, $c=\omega_1$ for $z_0=e_1$, $c=\omega_1+\omega_2$ for
$z_0=e_2$ and $c=\omega_2$ for $z_0=z_3$. 
Here 
\begin{equation}
\omega_1=\frac{K(\mathcal{K})}{\sqrt{e_1-e_3}} \ \ , \ \ 
\omega_2=i\frac{K(\mathcal{K'})}{\sqrt{e_1-e_3}} \ ,
\end{equation}
where $K$ denotes the complete elliptic integral of the first kind and $\mathcal{K}$ is the modulus of the elliptic integral with
\begin{eqnarray}\label{kelliptic}
\mathcal{K}&=&\sqrt{\frac{e_2-e_3}{e_1-e_3}}  \ .
\end{eqnarray}
Moreover we have $\mathcal{K'}$ = $\sqrt{1-\mathcal{K}^2}$.
For $D > 0$, $D=0$ and $D < 0$ there are various replacements of the Weierstrass
function $\wp$ by real valued Jacobi elliptic functions \cite{Ab}.

In the following, we will discuss different type of orbits, which we denote as follows~:
\begin{enumerate}

\item Bound orbit~: an orbit for which $r$ varies between two finite values.
\item Spherical orbit~: a special bound orbit for which $r$ is constant.
 \item Bound terminating orbit~: a bound orbit that ends at the singularity $r=0$.
\item Unbound terminating orbit~: an orbit for which $r$ varies between $r=0$ and infinity.
\item Escape orbit~: the orbit comes from infinity, approaches a finite $r$ and goes again
to $r=\infty$.
\end{enumerate}

\subsection{Geodesics for $D >0$ and $E^2-\varepsilon > 0$}
This case corresponds to the choice $V_{\rm eff}^{\rm min} < \frac{E^2-\varepsilon}{2} < V_{\rm eff}^{\rm max}$  and $E^2-\varepsilon > 0$ (region I in the $\lambda$-$\mu$-plot).
The effective potential and the corresponding characteristic polynomial $P(z)$ are shown
in Fig.\ref{potential_polynomial2}. The line $\frac{E^2-\varepsilon}{2}$ intersects the effective potential
$V_{\rm eff}(r)$ twice. The polynomial $P(z)$ has in fact three real zeros $e_1$, $e_2$, $e_3$.  However, for
$e_2 < z < e_3$ we have $P(z) < 0$ and in addition $e_1 < -1/3$.
We are hence allowed to integrate from $e_2$ to $z=-1/3$, which corresponds to integration from
$r=r_2$ to $r=\infty$ and from $e_3$ to $z=\infty$ which corresponds to integration from
$r=0$ to $r=r_3$.

\subsubsection{Escape orbit and light deflection}
\begin{figure}[htbp]
\begin{center}
\resizebox{6in}{!}{\includegraphics{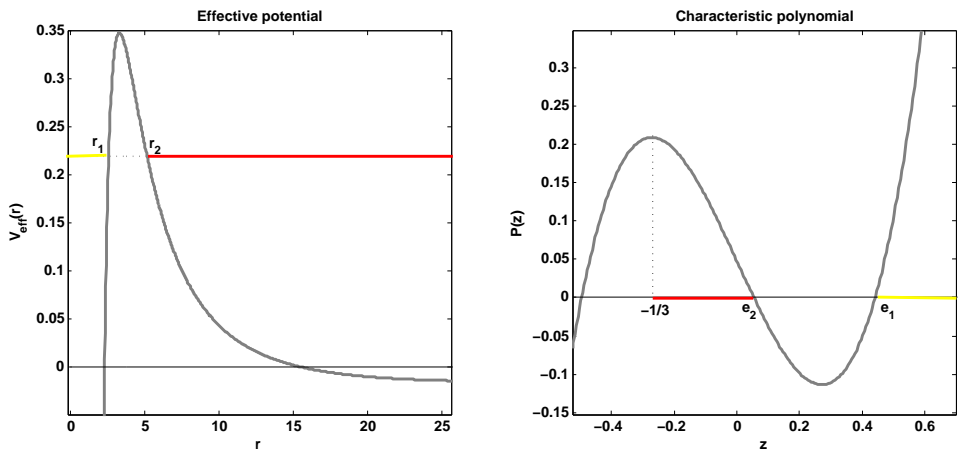}}
\end{center}
\caption{The effective potential $V_{\rm eff}(r)$ (left) and the characteristic polynomial $P(z)$ (right) of a massive test particle $\varepsilon =1$ for $D >0$. The red horizontal line corresponds to the
range of $r$ values for an escape orbit. \label{potential_polynomial2}}
\end{figure}

The range of allowed values for $r$ is indicated by the red horizontal line in Fig.\ref{potential_polynomial2} and is $r_2 \le r \le \infty$.

\begin{figure}[htbp]
\begin{center}
\subfigure[with $\beta$ = 0.33 ]{\includegraphics[scale=0.55]{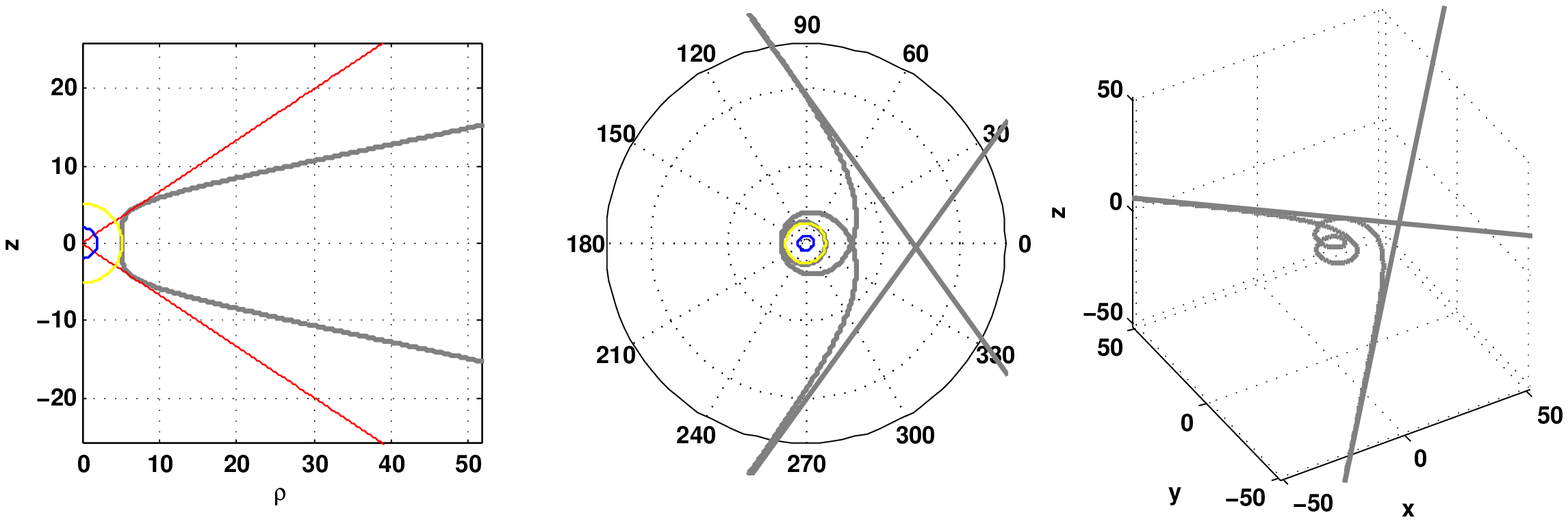}} \\
\subfigure[with $\beta$ = 0.50 ]{\includegraphics[scale=0.55]{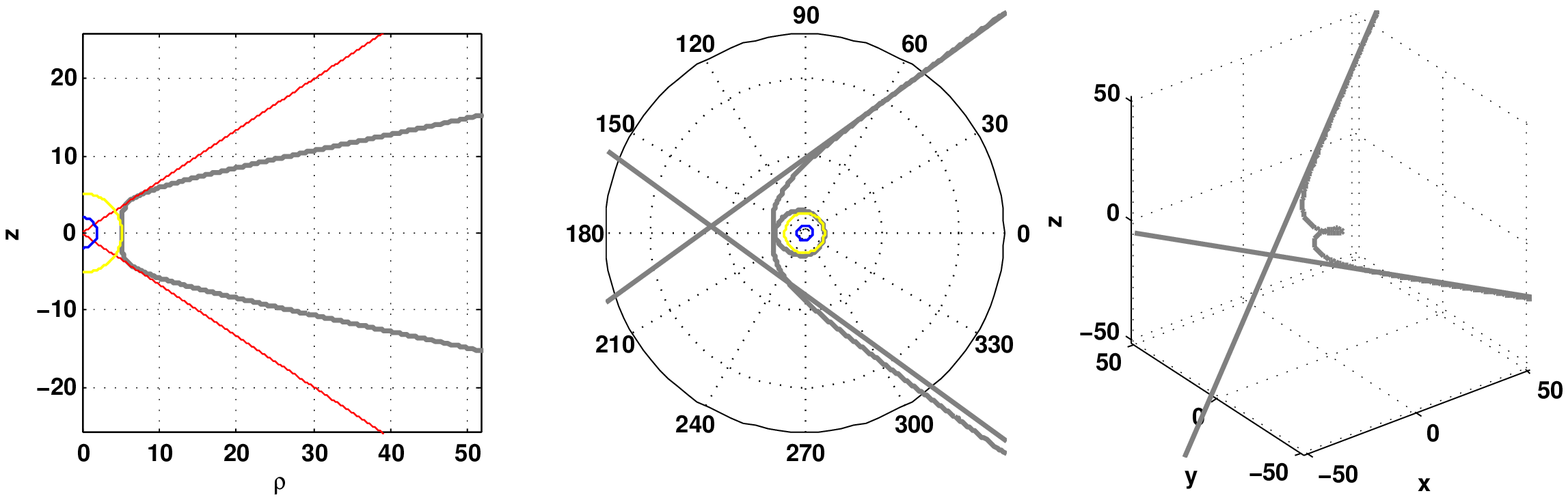}} \\
\subfigure[with $\beta$ = 0.99 ]{\includegraphics[scale=0.55]{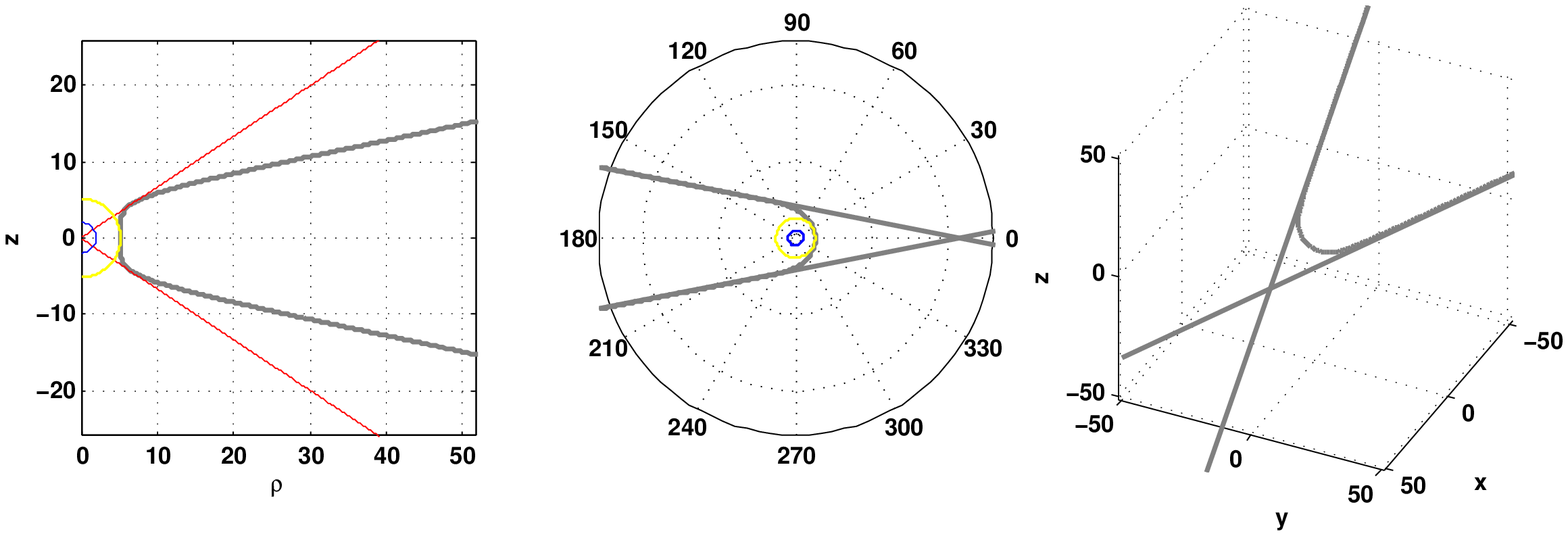}}
\end{center}
\caption{We show an escape orbit for a massive test particle ($\varepsilon=1$) in the $\rho$-$z$-plane
(with $\rho=r\sin\theta$, $z=r\cos\theta$) (left), in the plane perpendicular to $\vec{L}$ (middle) and in $\mathbb{R}^3$ (right).
(a), (b) and (c) correspond to $\beta=0.33$, $\beta=0.5$ and $\beta=0.99$, respectively.
Here, we have chosen $E=1.2$, $L_{z}=5$, $k=1.2$ and $m=1$. The yellow circles in the $\rho$-$z$-plane 
and in the plane perpendicular to $\vec{L}$ correspond to circles with radius $r_2$, respectively.}
\label{mslz5E105Esvb}
\end{figure}

\begin{figure}[htbp]
\begin{center}
\subfigure[with $\beta$ = 0.33]{\includegraphics[scale=0.55]{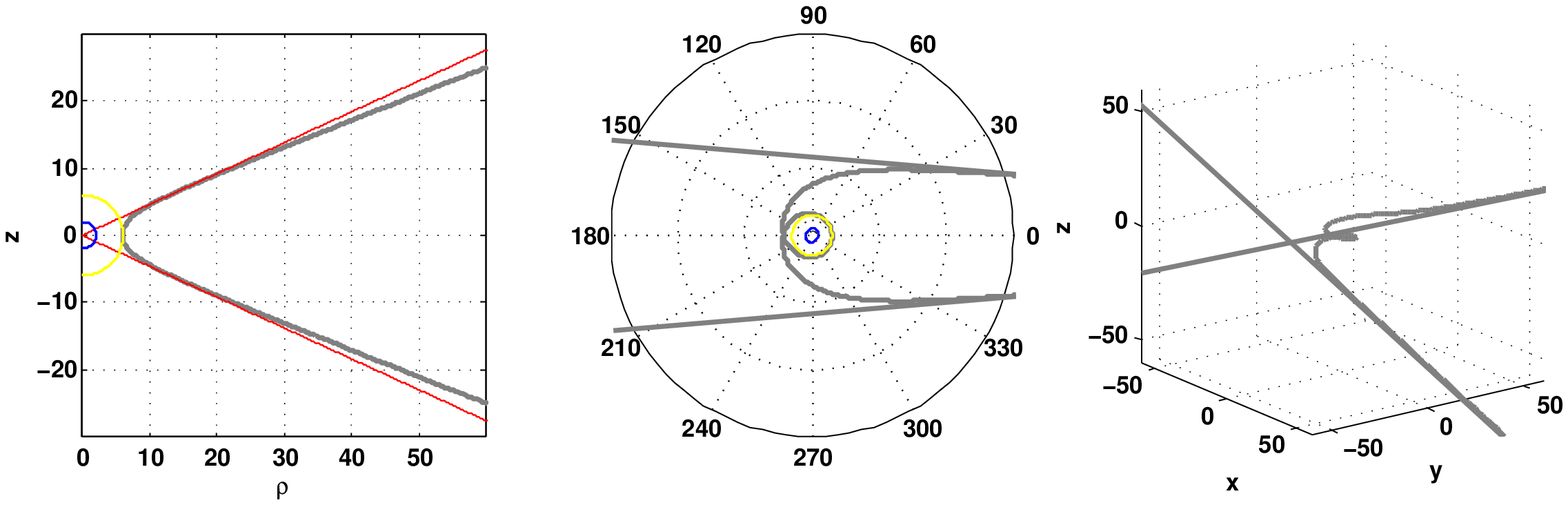}} \\
\subfigure[with $\beta$ = 0.50]{\includegraphics[scale=0.55]{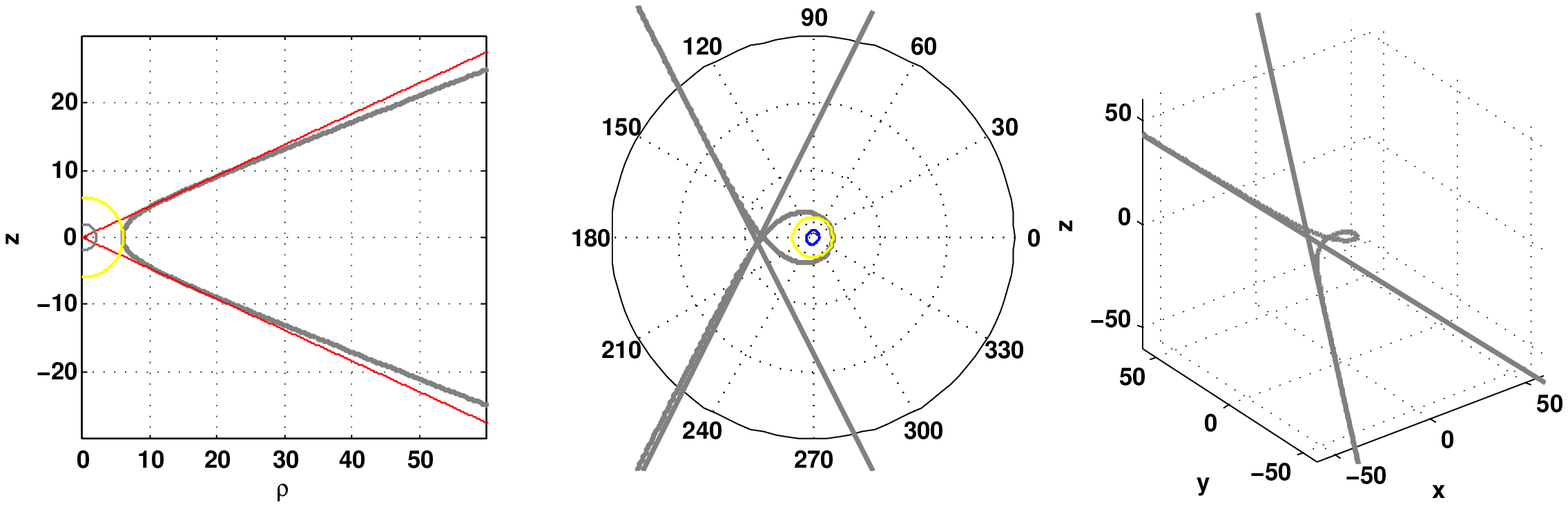}}\\
\subfigure[with $\beta$ = 0.96]{\includegraphics[scale=0.55]{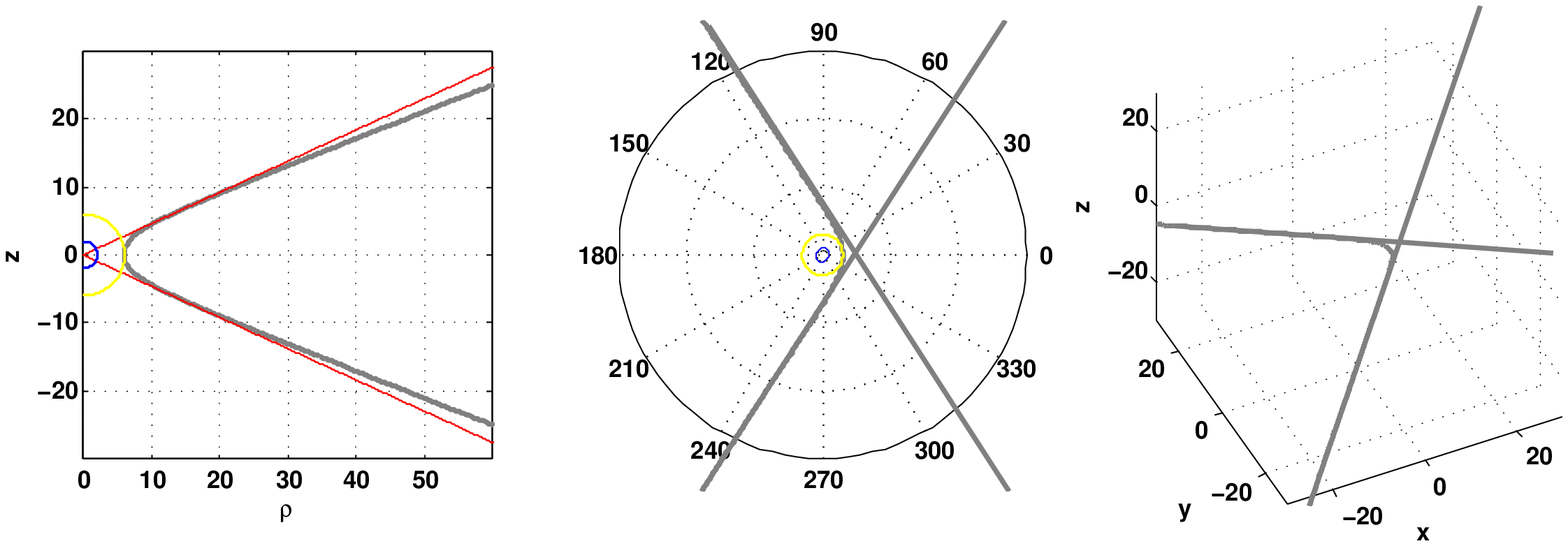}}\\
\end{center}
\caption{We show an escape orbit for a massless test particle ($\varepsilon=0$) in the $\rho$-$z$-plane
(with $\rho=r\sin\theta$, $z=r\cos\theta$) (left), in the plane perpendicular to $\vec{L}$ (middle) and in $\mathbb{R}^3$ (right).
(a), (b) and (c) correspond to $\beta=0.33$, $\beta=0.5$ and $\beta=0.96$, respectively.
Here, we have chosen $E=0.75$, $L_{z}=5$, $k=1.1$ and $m=1$. The yellow circles in the $\rho$-$z$-plane 
and in the plane perpendicular to $\vec{L}$ correspond to circles with radius $r_2$, respectively.}\label{mlposdiscE}
\end{figure}

The plots of escape orbits of a massive test particle and a massless test particle are given in Fig.\ref{mslz5E105Esvb} and Fig.\ref{mlposdiscE}, respectively. Obviously, for $\beta$ close to one, test particles
get simply deflected. However, if $\beta$ is significantly smaller than one, particles will approach
the minimal radius $r_2$, then whirl the black hole and finally move off to infinity again.
This is particularly important for the deflection of light. 

In order to see the effect of $\beta$ on the deflection of light ($\varepsilon=0$) we set $k=1$ (i.e. we choose $\theta=\pi/2$) and select the initial point at $r_0= r_{2}$. The requirement $r(\phi)=\infty$ imposes a bound on $\phi$ which reads:
\begin{eqnarray}
|\phi|\leq\frac{2}{\beta}\left[\frac{1}{\sqrt{e_1-e_3}}\int_{0}^{\varphi_{c}}\frac{d\varphi}{\sqrt{1-\mathcal{K}^{2}\sin^{2}(\varphi)}}+\omega_{1}\right]  \ .
\end{eqnarray}
Therefore the deflection angle $\Delta\phi$ is given by:
\begin{eqnarray}
\Delta\phi
=\frac{1}{\beta}\left[\frac{4}{\sqrt{e_1-e_3}}\int_{0}^{\varphi_{c}}\frac{d\varphi}{\sqrt{1-\mathcal{K}^{2}\sin^{2}(\varphi)}}+2\omega_{1}\right]+\pi\left(\frac{1}{\beta}-1\right)  \ .
\label{lighdeflec}
\end{eqnarray}
The term in the square bracket of (\ref{lighdeflec}) is the expression of $\Delta\phi$ for the ``pure''  Schwarzschild case, i.e. $\beta=1$ and the second term
takes care of the fact that the space--time has a conical deficit. 

Using the parametrized Post-Newtonian (PPN) formalism \cite{will1}, which describes deviations from
standard General Relativity, the angle of light deflection can be given by
 $\Delta\phi=\frac{1}{2}(1+\gamma_1) 1.75''=\frac{1}{2}(1+\gamma_1)(\Delta\phi)_S$ 
assuming that the parameter $\gamma_1$ is equal to unity for General Relativity. $(\Delta\phi)_S$ denotes the
General Relativity value assuming the exterior space-time of the massive body to be given by the Schwarzschild solution. 
Different experimental tests \cite{turyshev,wtb,bit} have given a value of $\gamma_1-1=(2.1\pm 2.3)\cdot 10^{-5}$, hence $\frac{\Delta\phi-(\Delta\phi)_S}{(\Delta\phi)_S} \lesssim 10^{-5}$.
If we now assume that
the deviations of $\Delta\phi$ from the Schwarzschild value are not due to a modification of General Relativity, but
due to the presence of a cosmic string, we can approximate the energy per unit length of the cosmic string.
Reinstalling factors of $c^2$ we then find
$(1-\beta)\lesssim 10^{-11}$ which for the deficit angle gives  $\Delta \lesssim 10^{-10}$. This transfers to a bound on the energy per unit length given by~: $\mu \lesssim 10^{16} \frac{\rm kg}{\rm m}$.



\subsubsection{Bound terminating orbit}
The range of allowed
$r$ values is indicated in Fig.\ref{potential_polynomial2} and is $0 \le r \le r_1$. We don't 
discuss this case in detail here, because it is qualitatively similar to the case with $D >0$ and $E^2-\varepsilon < 0$
(see discussion below).

\subsection{Geodesics for $D >0$ and $\frac{E^2-\varepsilon}{2} < 0$}
\begin{figure}[htbp]
\begin{center}
\resizebox{6in}{!}{\includegraphics{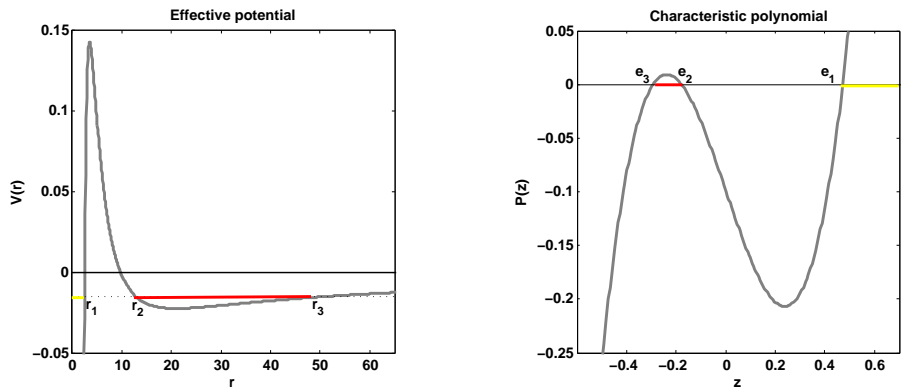}}
\end{center}
\caption{The effective potential $V_{\rm eff}(r)$ (left) and the characteristic polynomial $P(z)$ (right) of a massive test particle $\varepsilon =1$ for $D >0$. The red horizontal line corresponds to the
range of $r$ values for a bound orbit, while the yellow line corresponds to the range of $r$ for a bound terminating orbit. \label{potential_polynomial}}
\end{figure}

This case corresponds to the choice $V_{\rm eff}^{\rm min} < \frac{E^2-\varepsilon}{2} < V_{\rm eff}^{\rm max}$  and $E^2-\varepsilon < 0$ (region II in the $\lambda$-$\mu$-plot).
The effective potential and the corresponding characteristic polynomial $P(z)$ are shown
in Fig. \ref{potential_polynomial}. The line $\frac{E^2-\varepsilon}{2}$ intersects the effective potential
$V_{\rm eff}(r)$ three times at $r_1$, $r_2$ and $r_3$. These intersection points correspond to the zeros of
the characteristic polynomial, which we denote by $e_1$, $e_2$ and $e_3$. Apparently, we are allowed
to integrate between $\infty$ and $e_1$, which corresponds to integration from $r=0$ to $r=r_1$,
and from $e_2$ to $e_3$, which corresponds to integration from $r=r_2$ to $r=r_3$.

\subsubsection{Bound terminating orbit}
The range of allowed $r$ values is indicated by the yellow line in Fig.\ref{potential_polynomial} and is $0 \le r \le r_1$.

\begin{figure}[htbp]
\begin{center}
\subfigure[with $\beta$ = 0.25]{\includegraphics[scale=0.55]{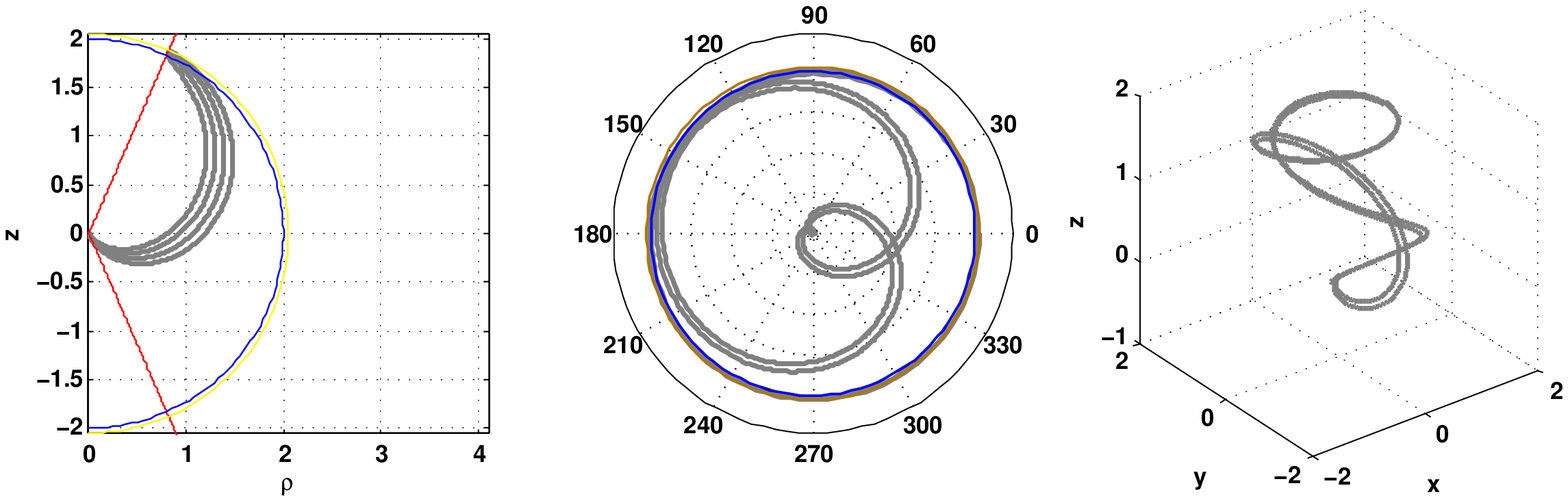}} \\
\subfigure[with $\beta$ = 0.75]{\includegraphics[scale=0.55]{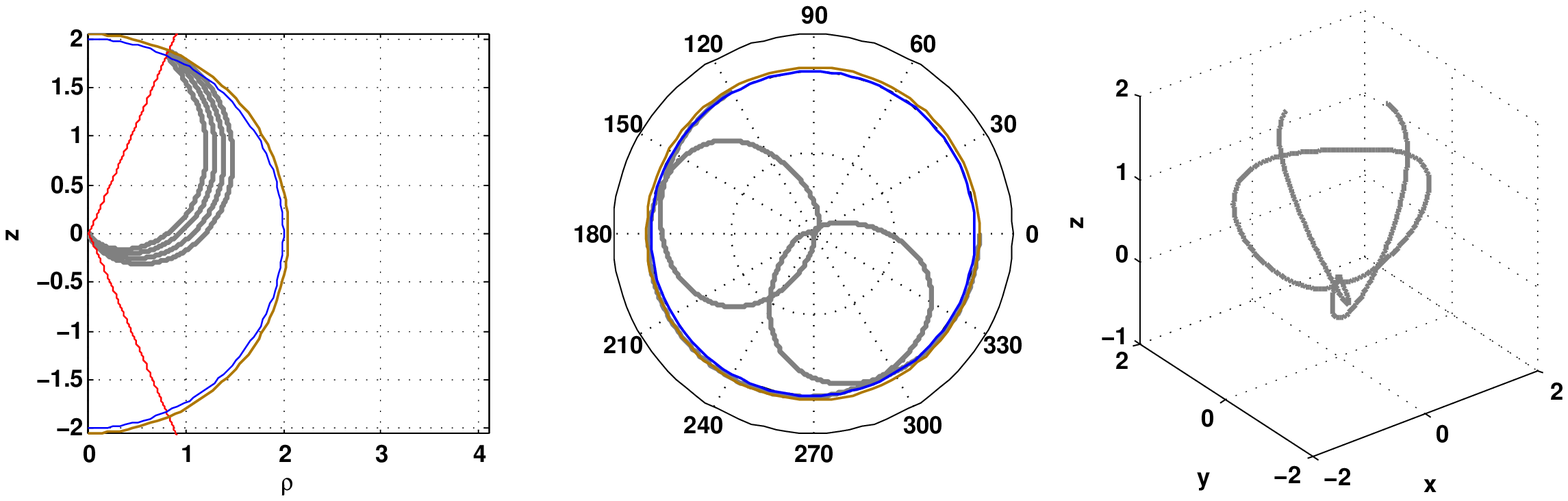}} \\
\subfigure[with $\beta$ = 0.99]{\includegraphics[scale=0.55]{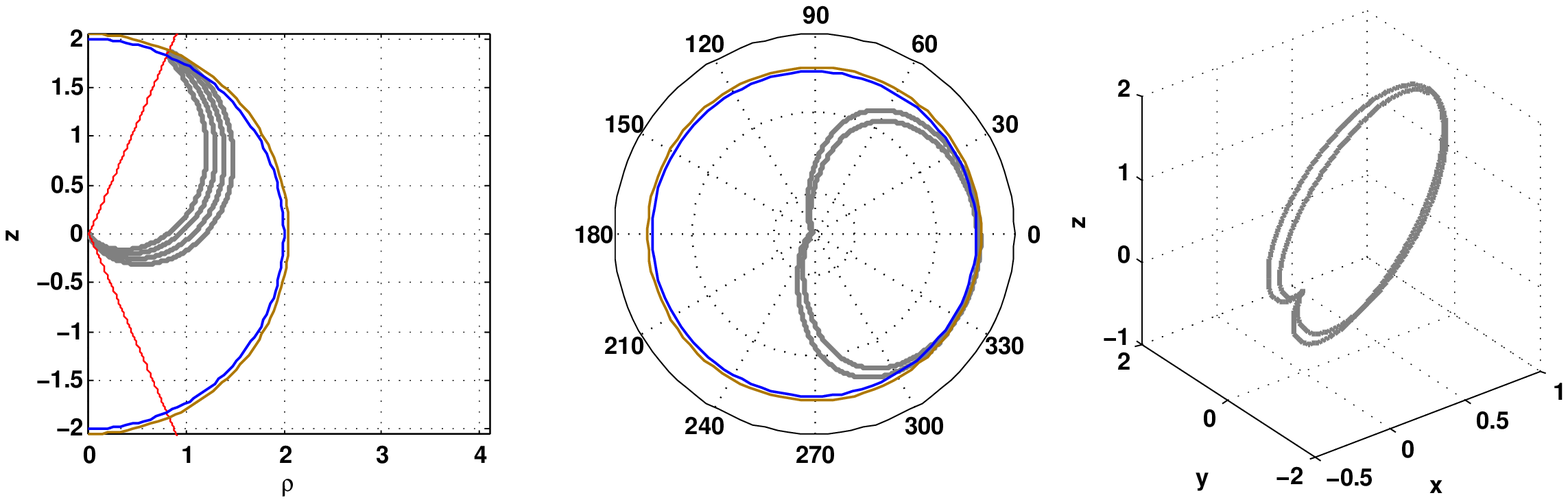}}
\end{center}
\caption{We show a bound terminating orbit for a massive test particle ($\varepsilon=1$) in the $\rho$-$z$-plane
(with $\rho=r\sin\theta$, $z=r\cos\theta$) (left), in the $x$-$y$-plane (middle) and in $\mathbb{R}^3$ (right).
(a), (b) and (c) correspond to $\beta=0.25$, $\beta=0.75$ and $\beta=0.99$, respectively.
Here, we have chosen $E=0.9975$, $L_{z}=5$, $k=2.5$ and $m=1$. The blue and the dark yellow circles in the $\rho$-$z$-plane 
and in the $x$-$y$-plane correspond to circles with Schwarzschild radius $r_{s}=2m=2$and maximal radius $r_1$, respectively. }
\label{mslz5E998BTvb}
\end{figure}

A plot of a bound terminating orbit for a massive test particle is given in Fig. \ref{mslz5E998BTvb}. 
We emphasize on the deformation of the orbits
when changing the deficit parameter $\beta$. While for $\beta=0.99$, the orbit is still nearly planar, this changes
with the decrease of $\beta$, i.e. the increase of the deficit angle. While the orbit looks
heart-shaped for $\beta=0.99$, the increase of the deficit angle leads to a pretzel-like structure.

\subsubsection{Bound orbit and perihelion shift}
The range of allowed
$r$-values is indicated by the red line in Fig.\ref{potential_polynomial} and is $r_2 \le r \le r_3$.

\begin{figure}[htbp]
\begin{center}
\subfigure[with $\beta$ = 0.25]{\includegraphics[scale=0.55]{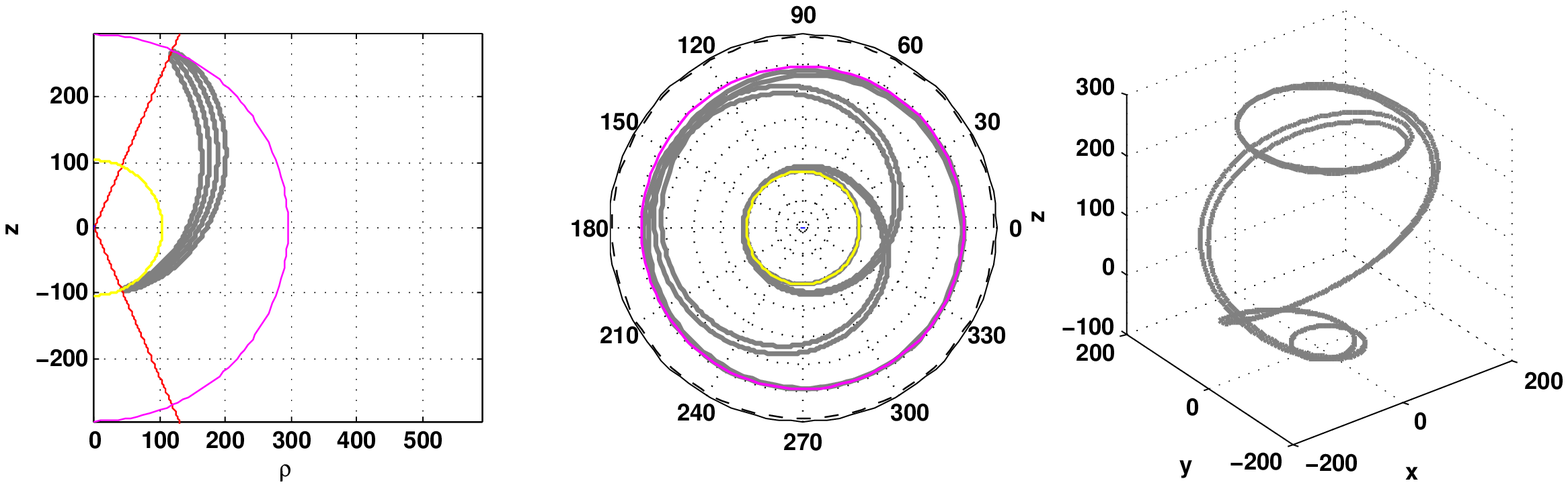}} \\
\subfigure[with $\beta$ = 0.75]{\includegraphics[scale=0.55]{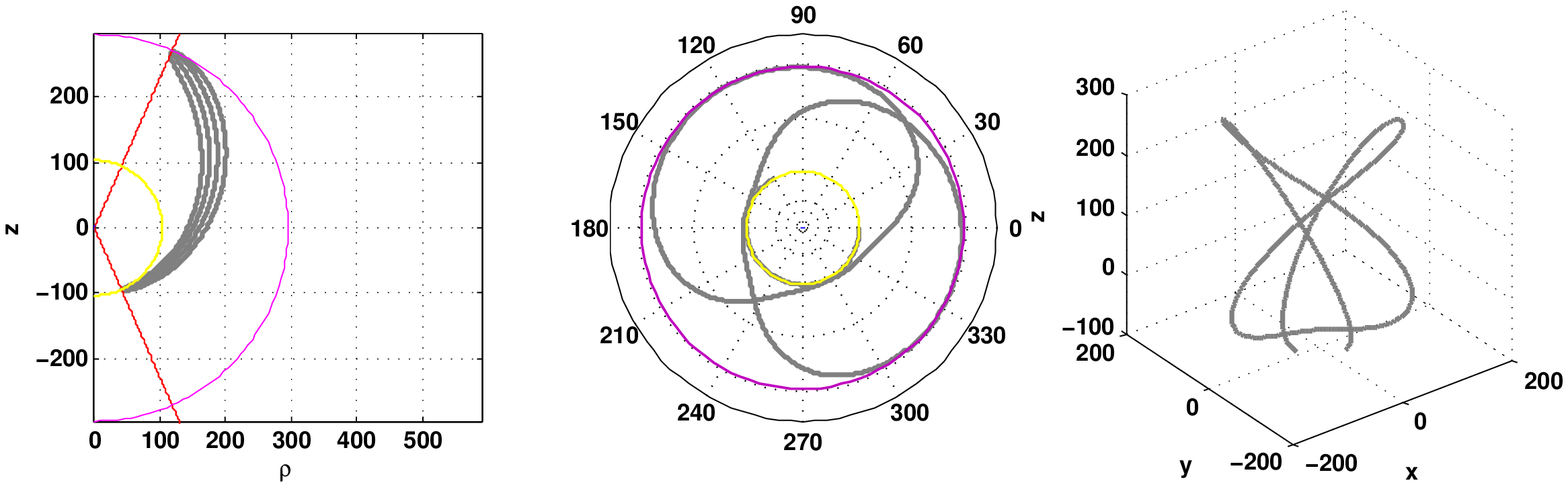}} \\
\subfigure[with $\beta$ = 0.99]{\includegraphics[scale=0.55]{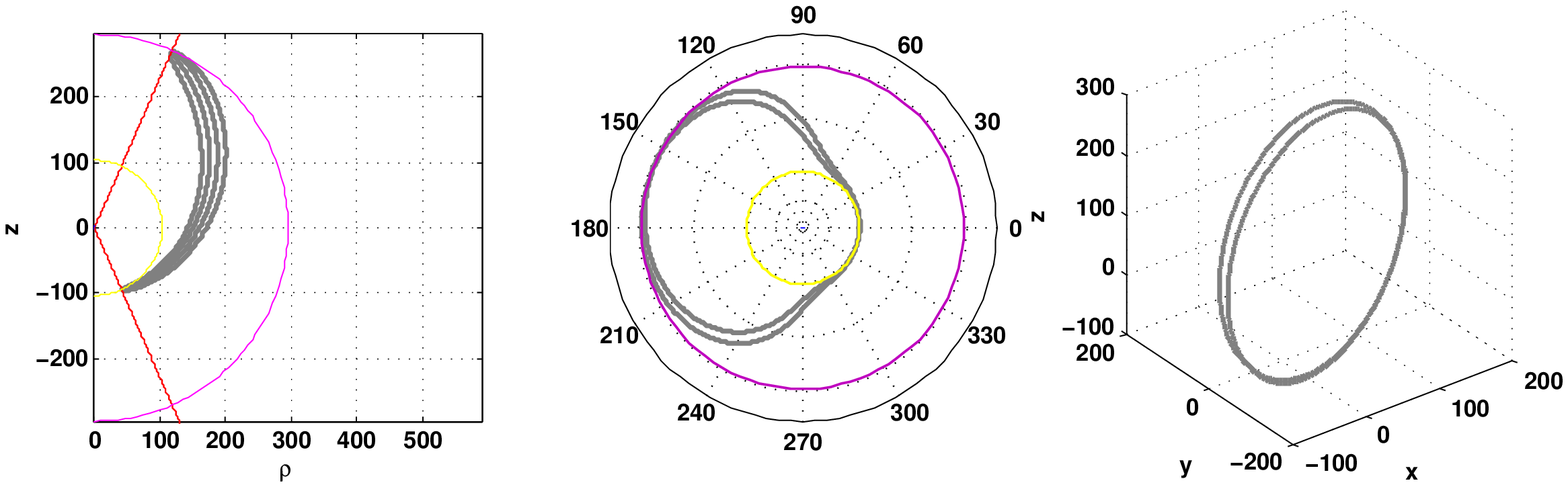}}
\end{center}
\caption{We show a bound orbit for a massive test particle ($\varepsilon=1$) in the $\rho$-$z$-plane
(with $\rho=r\sin\theta$, $z=r\cos\theta$) (left), in the plane perpendicular to $\vec{L}$ (middle) and in $\mathbb{R}^3$ (right).
(a), (b) and (c) correspond to $\beta=0.25$, $\beta=0.75$ and $\beta=0.99$, respectively.
Here, we have chosen $E=0.9975$, $L_{z}=5$, $k=2.5$ and $m=1$. The yellow and magenta circles in the $\rho$-$z$-plane 
and in the plane perpendicular to $\vec{L}$ correspond to circles with radius $r_2$ and maximal radius $r_3$, respectively.}
\label{mslz5E998Bvb}
\end{figure}

A bound orbit of a massive test particle is given in Fig. \ref{mslz5E998Bvb} for different choices of $\beta$.
For $\beta$ close to one, the orbit corresponds to a standard closed orbit, while decreasing $\beta$ leads to the appearance of additional loops in the orbit (see the cases $\beta=0.25$ and $\beta=0.75$). This is important
in order to understand the perihelion shift of test particles orbiting a massive body that is pierced
by a cosmic string.

For bound orbits the perihelion shift depends on the deficit
parameter $\beta$. This effect has been calculated approximately in \cite{fbl}. Here, we will give the exact analytic formula
for the perihelion shift. We choose $k = 1$ in order to study planar orbits with $\theta=\pi/2$. We then have
for the motion in the plane:
\begin{eqnarray}
r(\phi)=\frac{2m}{\wp(\frac{\beta\phi}{2})+\frac{1}{3}}  \ .
\end{eqnarray}
When the test particle moves from $r_{3}\rightarrow r_2 \rightarrow r_3$ we find for the change in $\phi$:

\begin{eqnarray}
\delta\phi&=&\frac{4}{\beta}\frac{K(\mathcal{K})}{\sqrt{e_{1}-e_{3}}}
\end{eqnarray}
such that the perihelion shift $\Delta\phi$ reads
\begin{eqnarray}
\Delta\phi=\delta\phi-2\pi=\frac{4}{\beta}\frac{K(\mathcal{K})}{\sqrt{e_{1}-e_{3}}} -2\pi  \ .
\end{eqnarray}

\begin{figure}[htbp]
\begin{center}
\resizebox{!}{2in}{\includegraphics{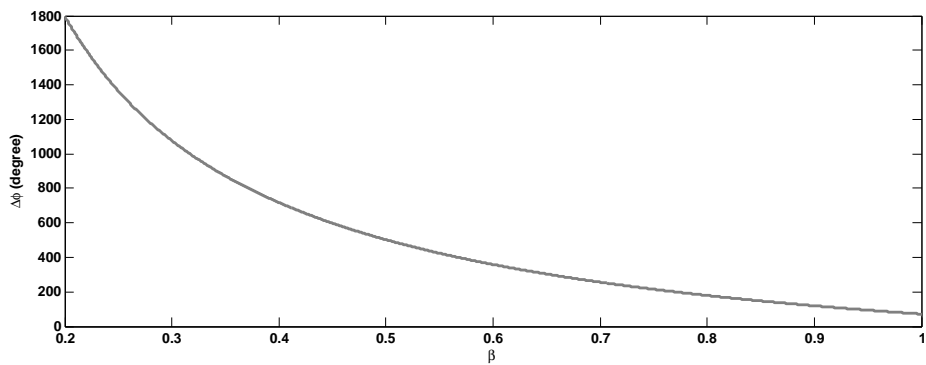}}
\end{center}
\caption{The perihelion shift $\Delta\phi$ in dependence on the deficit parameter $\beta$ for
$m=1$, $L_z=5$, $k=1$ and $E=0.999$.}\label{deficit_shift}
\end{figure}

In Fig.\ref{deficit_shift}, we plot the value of $\Delta\phi$ in dependence on $\beta$ for $m=1$, $L_z=5$, $k=1$ and $E=0.999$.
Obviously, the perihelion shift can become quite large due to the conical nature of the space-time.
Perihelion shifts in the solar system are quite small and the level of agreement of the measured
value with that predicted by assuming the gravitational field of the sun to be described
by the Schwarzschild solution can be used  to give an upper bound on the energy density per unit length of
a cosmic string present in the solar system. 
However, a large perihelion shift of $39\,^{\circ}$ per orbit has been observed in a massive binary black hole system \cite{valtonen}. Note that if we would do a similar
approximation as that done in \cite{fbl}, we would find 
\begin{equation}
 \Delta\phi=\frac{1}{\beta} (\Delta\phi)_S + 2\pi\left(\frac{1}{\beta}-1\right)  \ ,
\end{equation}
where $(\Delta\phi)_S$ denotes the perihelion shift in the Schwarzschild case $\beta=1$. Within the PPN formalism \cite{will1}, the perihelion shift can be written as
$\Delta\phi=42.98''\left(\frac{1}{3}(2+2\gamma_1-\gamma_2)\right)$
assuming that the parameters $\gamma_1$ and $\gamma_2$ are equal to unity for General Relativity
and that the quadrupole moment of the sun vanishes. Different experimental tests \cite{turyshev,wtb,bit} have given
$\gamma_2-1=(1.2\pm 1.1)\cdot 10^{-4}$, hence $\frac{\Delta\phi-(\Delta\phi)_S}{(\Delta\phi)_S} \lesssim 10^{-4}$.

If we now assume that
the deviations of $\Delta\phi$ from the Schwarzschild value are not due to a modification of General Relativity, but
due to the presence of a cosmic string, we can approximate the energy per unit length of the cosmic string.
Again reinstalling factors of $c^2$ we find 
$(1-\beta)\lesssim 10^{-10}$ which for the deficit angle gives $\Delta \lesssim 10^{-9}$. This transfers to
a bound on the  energy per unit length which is given by~: $\mu \lesssim 10^{17} \frac{\rm kg}{\rm m}$.

\subsection{Geodesics for $D <0$ and $E^2 -\frac{2}{3}\varepsilon > \frac{L^2}{54m^2}$}
\begin{figure}[htbp]
\begin{center}
\resizebox{6in}{!}{\includegraphics{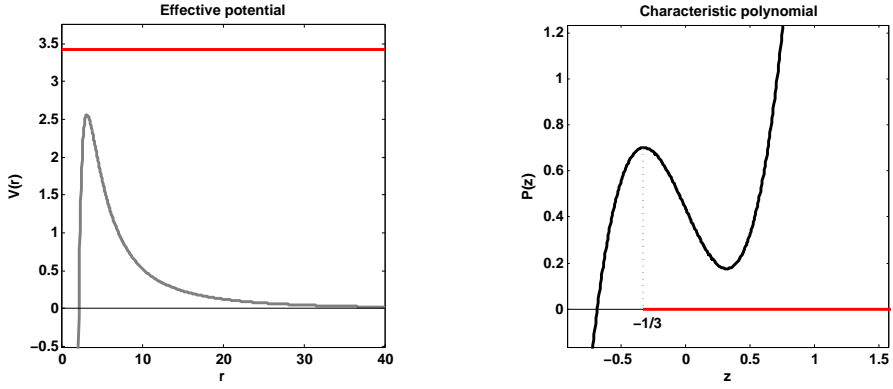}}
\end{center}
\caption{The effective potential $V_{\rm eff}(r)$ (left) and the characteristic polynomial $P(z)$ (right) for $D < 0$
and $E^2-\frac{2}{3}\varepsilon > \frac{L^2}{54m^2}$. The red horizontal line corresponds to the
range of $r$ values for an escape orbit.\label{potential_polynomial4}}
\end{figure}

This case corresponds to the choice $V_{\rm eff}^{\rm max} < \frac{E^2-\varepsilon}{2}$ (region III in the $\lambda$-$\mu$-plot).
The effective potential and the corresponding characteristic polynomial $P(z)$ are shown
in Fig. \ref{potential_polynomial4}. The line $\frac{E^2-\varepsilon}{2}$ does not intersect the effective potential
$V_{\rm eff}(r)$. The polynomial $P(z)$ has a single real zero at $e_1$, however $e_1 < -1/3$. We are allowed
to integrate between $z=\infty$ and $z=-1/3$, which corresponds to integration from $r=0$ to $r=\infty$.

\subsubsection{Escape orbit}
\begin{figure}[htbp]
\begin{center}
\resizebox{6in}{!}{\includegraphics{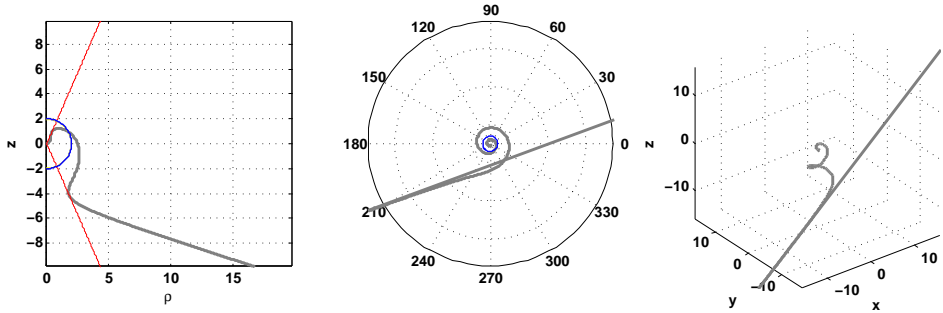}}
\end{center}
\caption{We show a bound terminating orbit for a massive test particle ($\varepsilon=1$) in the $\rho$-$z$-plane
(with $\rho=r\sin\theta$, $z=r\cos\theta$) (left), in the plane perpendicular to $\vec{L}$ (middle) and in $\mathbb{R}^3$ (right).
Here, we have chosen $\beta=0.4$, $E=2.8$, $L_{z}=5$, $k=2.5$ and $m=1$. The blue circles in the $\rho$-$z$-plane 
and in the plane perpendicular to $\vec{L}$ correspond to circles with Schwarzschild radius $r_s=2m=2$.}\label{msnegdiscE280b040lz5e}
\end{figure}

The range of allowed $r$ values is indicated by the red horizontal line in Fig.\ref{potential_polynomial4}
and corresponds to $0 \le r \le \infty$. 
The plot of an escape orbit in this case is shown in Fig.\ref{msnegdiscE280b040lz5e}. The particle
comes in from infinity and -- after a few loops -- ends in $r=0$.

\subsection{Geodesics for $D <0$ and $E^2 -\frac{2}{3}\varepsilon < \frac{L^2}{54m^2}$}
This case corresponds to the choice $V_{\rm eff}^{\rm min} > \frac{E^2-\varepsilon}{2}$ (region IV in the $\lambda$-$\mu$-plot).
The effective potential and the corresponding characteristic polynomial $P(z)$ are shown
in Fig. \ref{potential_polynomial3}. The line $\frac{E^2-\varepsilon}{2}$ intersects the effective potential
$V_{\rm eff}(r)$ only once at $r_1$. This intersection point corresponds to the single real zero of
the characteristic polynomial, which we denote by $e_1$. Apparently, we are allowed
to integrate between $\infty$ and $e_1$, which corresponds to integration from $r=0$ to $r=r_1$.

\subsubsection{Bound terminating orbits}

\begin{figure}[htbp]
\begin{center}
\resizebox{6in}{!}{\includegraphics{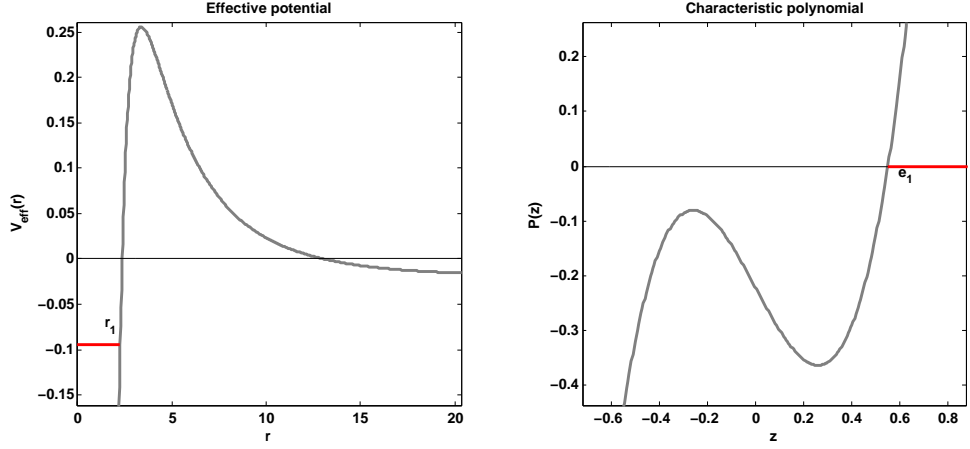}}
\end{center}
\caption{The effective potential $V_{\rm eff}(r)$ (left) and the characteristic polynomial $P(z)$ (right) for
$D < 0$ and $E^2-\frac{2}{3}\varepsilon < \frac{L^2}{54m^2}$. The red horizontal line corresponds to the
range of $r$ values for a bound terminating orbit. \label{potential_polynomial3}}
\end{figure}

The range of allowed $r$ values is indicated by the red line in Fig.\ref{potential_polynomial3} and is $0 \le r \le r_1$.

\begin{figure}[htbp]
\begin{center}
\resizebox{6in}{!}{\includegraphics{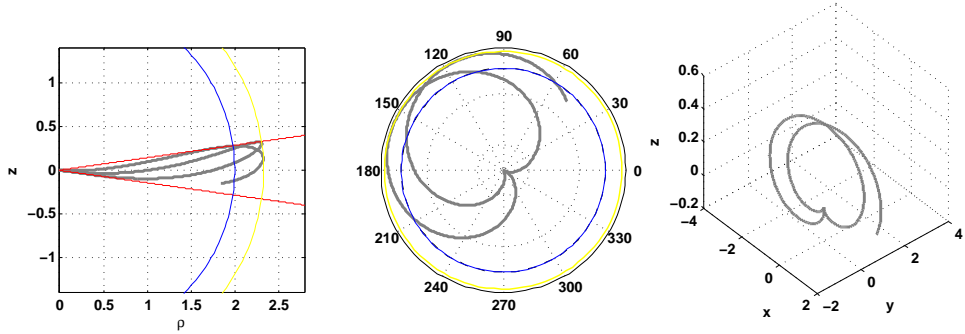}}
\end{center}
\caption{We show a bound terminating orbit for a massive test particle ($\varepsilon=1$) in the $\rho$-$z$-plane
(with $\rho=r\sin\theta$, $z=r\cos\theta$) (left), in the plane perpendicular to $\vec{L}$ (middle) and in $\mathbb{R}^3$ (right).
Here, we have chosen $\beta=0.99$, $E=0.9$, $L_{z}=5$, $1/k=0.99$ and $m=1$. The dark yellow and blue circles in the $\rho$-$z$-plane 
and in the plane perpendicular to $\vec{L}$ correspond to circles with radius $r_1$ and Schwarzschild radius $r_s=2m=2$, respectively.}\label{msnegdiscE090b099lz5bt}
\end{figure}

The example of a bound terminating orbit for a massive test particle for $\beta=0.99$ is given in Fig.\ref{msnegdiscE090b099lz5bt}.

\subsection{Geodesics for $D =0$ and $E^2 -\varepsilon > 0$}
\begin{figure}[htbp]
\begin{center}
\resizebox{6in}{!}{\includegraphics{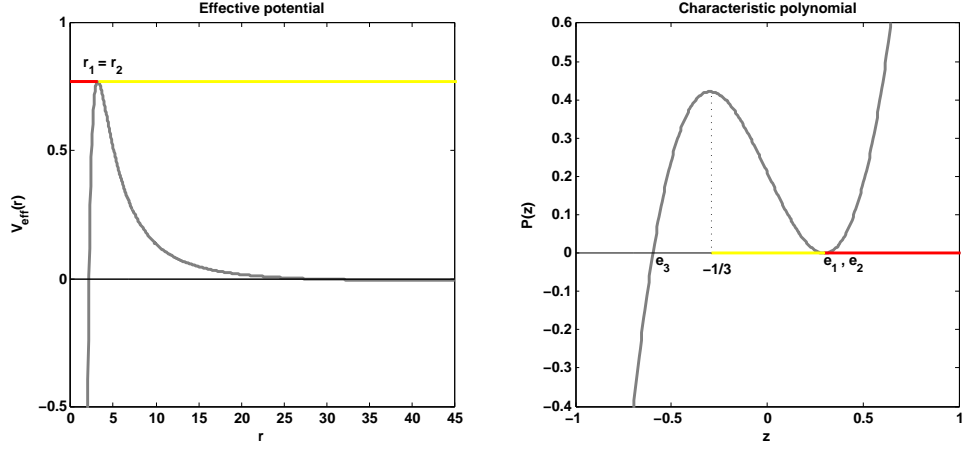}}
\end{center}
\caption{The effective potential $V_{\rm eff}(r)$ (left) and the characteristic polynomial $P(z)$ (right) for $D =0$
and $E^2-\varepsilon > 0$. The red horizontal line corresponds to the
range of $r$ values for a bound terminating orbit, while the yellow line corresponds to the range of $r$ for an unbound terminating orbit. \label{potential_polynomial5}}
\end{figure}

This case corresponds to the choice $V_{\rm eff}^{\rm max} = \frac{E^2-\varepsilon}{2} >0$.
The effective potential and the corresponding characteristic polynomial $P(z)$ are shown
in Fig. \ref{potential_polynomial5}. The line $\frac{E^2-\varepsilon}{2}$ intersects the effective potential
$V_{\rm eff}(r)$ only once at $r_1=r_2$, i.e. at the maximum of the effective potential.
 This intersection point corresponds to a double zero of
the characteristic polynomial, which we denote by $e_1=e_2$. Apparently, we are allowed
to integrate between $\infty$ and $e_1$, which corresponds to integration from $r=0$ to $r=r_1$ and from
$e_1$ to $z=-1/3$, which corresponds to integration from $r=r_1$ to $r=\infty$.

Note that the spherical orbit with $r=r_1=r_2=const.$ corresponds to the unstable
spherical orbit discussed in \cite{gm}.

\subsubsection{Unbound terminating orbit}
The range of allowed $r$ values is indicated by the yellow
line in (\ref{potential_polynomial5}) and is $r_1\le r \le \infty$.

\begin{figure}[htbp]
\begin{center}
\resizebox{6in}{!}{\includegraphics{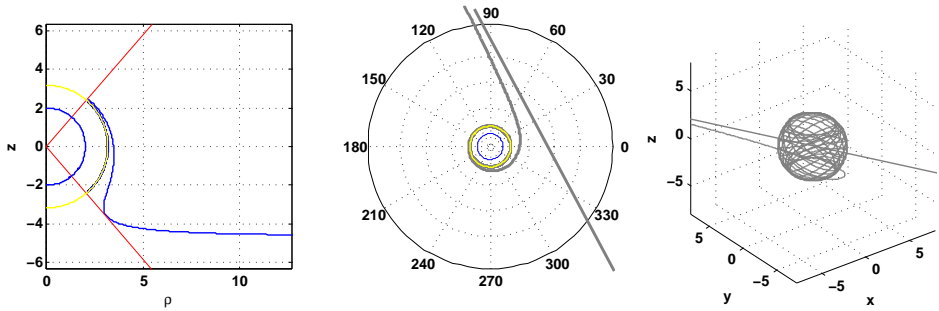}}
\caption{We show an unbound terminating orbit for a massive test particle ($\varepsilon=1$) in the $\rho$-$z$-plane
(with $\rho=r\sin\theta$, $z=r\cos\theta$) (left), in the plane perpendicular to $\vec{L}$ (middle) and in $\mathbb{R}^3$ (right).
We have chosen $\beta=0.65$, $E=1.594$, $L_{z}=5$, $1/k=0.65$ and $m=1$. The dark yellow circles in the $\rho$-$z$-plane 
and in the plane perpendicular to $\vec{L}$ correspond to circles with radius $r_1=r_2$.}\label{mszrdiscb065E159Lz5UB}
\end{center}
\end{figure}

An example of an unbound terminating orbit is given in Fig.\ref{mszrdiscb065E159Lz5UB} for $\beta=0.65$. The test particle
comes from infinity and approaches an unstable spherical orbit with $r=r_1=r_2=const.$. 

\subsubsection{Bound terminating orbit}
In this particular case the effective
potential $V_{\rm eff}(r)$ has its maximum at the maximal radius $r=r_1$ of the orbit. The range of allowed $r$ values is indicated by the red
line in (\ref{potential_polynomial5}) and is $0\le r \le r_1$.

\begin{figure}[htbp]
\begin{center}
\resizebox{6in}{!}{\includegraphics{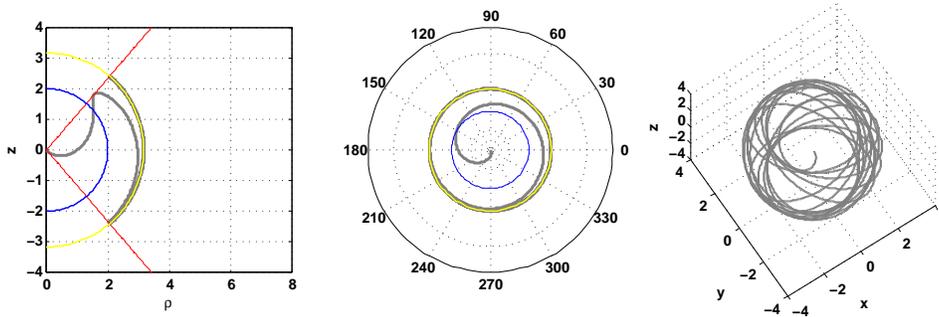}}
\caption{We show a bound terminating orbit for a massive test particle ($\varepsilon=1$) in the $\rho$-$z$-plane
(with $\rho=r\sin\theta$, $z=r\cos\theta$) (left), in the plane perpendicular to $\vec{L}$ (middle) and in $\mathbb{R}^3$ (right).
We have chosen $\beta=0.65$, $E=1.594$, $L_{z}=5$, $1/k=0.65$ and $m=1$. The dark yellow and blue circles in the $\rho$-$z$-plane 
and in the plane perpendicular to $\vec{L}$ correspond to circles with radius $r_1$ and Schwarzschild radius $r_s=2m=2$, respectively. }\label{mszrdiscb065E159Lz5}
\end{center}
\end{figure}

An example of a bound terminating orbit is shown in Fig.\ref{mszrdiscb065E159Lz5} for $\beta=0.65$.
The particle starts at $r=0$ and approaches the unstable spherical orbit at $r=r_1=r_2=const.$.

\subsection{Geodesics for $D =0$ and $E^2 -\varepsilon < 0$}
\begin{figure}[htbp]
\begin{center}
\resizebox{6in}{!}{\includegraphics{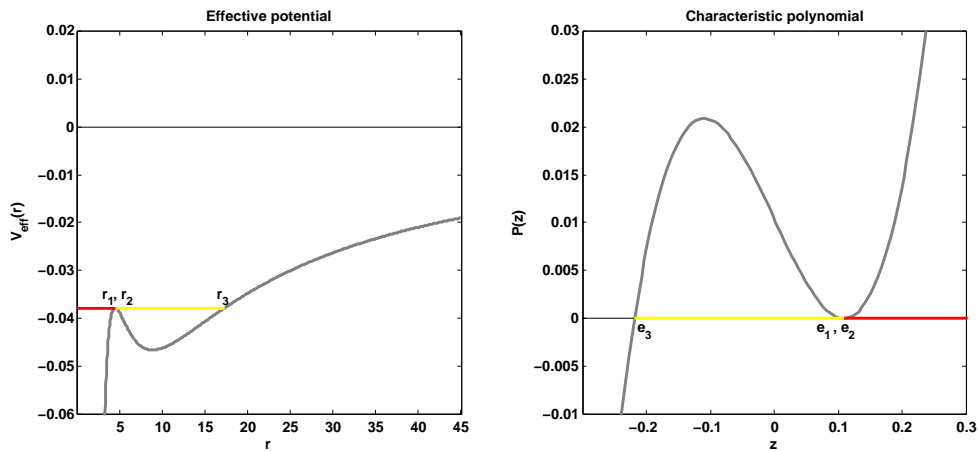}}
\caption{The effective potential $V_{\rm eff}(r)$ (left) and the characteristic polynomial $P(z)$ (right) for $D=0$
and $E^2-\varepsilon< 0$. The red horizontal line corresponds to the
range of $r$ values for a bound terminating orbit, while the yellow line corresponds to the range of $r$ for a bound orbit. \label{potential_polynomial6}}
\end{center}
\end{figure}

This case corresponds to the choice $V_{\rm eff}^{\rm max} = \frac{E^2-\varepsilon}{2} < 0$.
The effective potential and the corresponding characteristic polynomial $P(z)$ are shown
in Fig. \ref{potential_polynomial6}. The line $\frac{E^2-\varepsilon}{2}$ intersects the effective potential
$V_{\rm eff}(r)$  at $r_1=r_2$ and at $r_3$.
 These intersection points correspond to a double zero $e_1=e_2$ and a simple zero at $e_3$.
We are only allowed to 
integrate between $e_1=e_2$ and $e_3$ which corresponds to integration between $r_1=r_2$ and $r_3$ as well
as between $z=\infty$ and $e_3$ which corresponds to integration between $r=0$ and $r_1=r_2$.

\subsubsection{Bound orbit}
The range of allowed $r$ values
is indicated in Fig.\ref{potential_polynomial6} by the horizontal yellow line and is  $r_1 \le r \le r_3$.

\begin{figure}[htbp]
\begin{center}
\resizebox{6in}{!}{\includegraphics{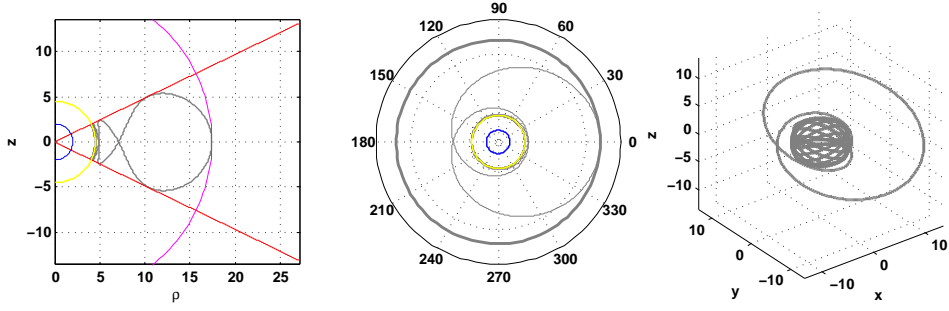}}
\caption{We show a bound orbit for a massive test particle ($\varepsilon=1$) in the $\rho$-$z$-plane
(with $\rho=r\sin\theta$, $z=r\cos\theta$) (left), in the plane perpendicular to $\vec{L}$ (middle) and in $\mathbb{R}^3$ (right).
We have chosen $\beta=0.9$, $E=0.9614$, $L_{z}=3.3$, $k=\frac{10}{9}$ and $m=1$. The dark yellow and violet circles in the $\rho$-$z$-plane 
and in the plane perpendicular to $\vec{L}$ correspond to circles with radius $r_2$ and $r_3$, respectively.}\label{mszrdiscb090E0961Lz33}
\end{center}
\end{figure}

An example of a bound orbit is given in Fig.\ref{mszrdiscb090E0961Lz33} for $\beta=0.9$. This is an example
of a so-called homoclinic orbit. Homoclinic orbits are orbits that approach unstable circular orbits and have been
discussed extensively in the context of gravitational wave production in binary system.
There they play the role of a transition between the late inspiral and the final plunge \cite{loc}.

\subsubsection{Bound terminating orbit}
\begin{figure}[htbp]
\begin{center}
\resizebox{6in}{!}{\includegraphics{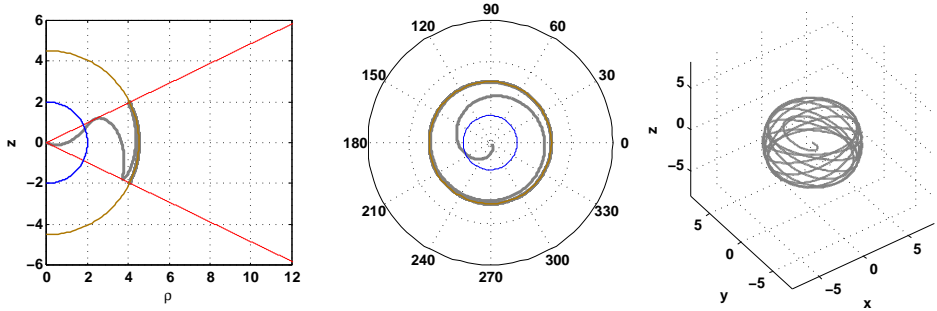}}
\caption{We show a bound terminating orbit for a massive test particle ($\varepsilon=1$) in the $\rho$-$z$-plane
(with $\rho=r\sin\theta$, $z=r\cos\theta$) (left), in the plane perpendicular to $\vec{L}$ (middle) and in $\mathbb{R}^3$ (right).
We have chosen $\beta=0.9$, $E=0.9614$, $L_{z}=3.3$, $k=\frac{10}{9}$ and $m=1$. The dark yellow circles in the $\rho$-$z$-plane 
and in the plane perpendicular to $\vec{L}$ correspond to circles with radius $r_2$.}\label{mszrdiscb090E0961Lz33bt}
\end{center}
\end{figure}

In this particular case the effective potential has its maximum at the maximal radius $r=r_1$ of the orbit. 
An example of a bound terminating orbit is given in Fig.\ref{mszrdiscb090E0961Lz33bt}.

\clearpage

\subsection{Geodesics for $D=0$ and $\frac{E^2}{L^2}=\frac{1}{54 m^2}$}
\begin{figure}[htbp]
\begin{center}
\resizebox{6in}{!}{\includegraphics{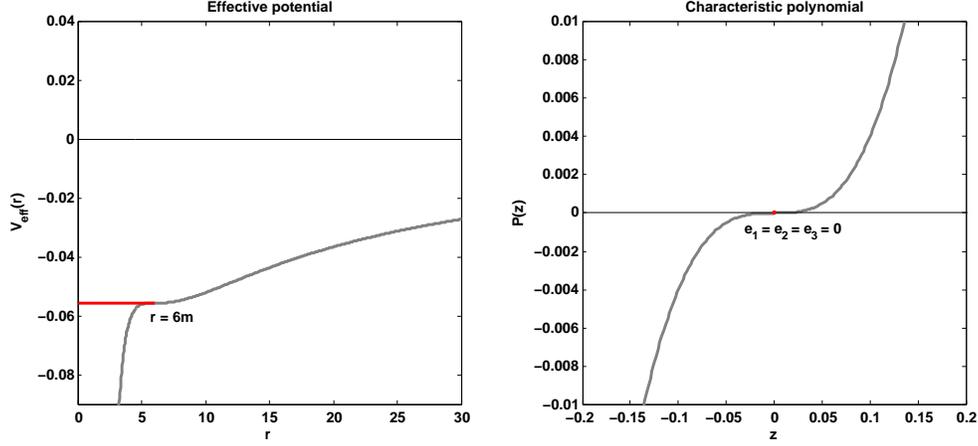}}
\caption{The effective potential $V_{\rm eff}(r)$ (left) and the characteristic polynomial $P(z)$ (right) for $D=0$
and $\frac{E^2}{L^2}= \frac{1}{54m^2}$. The red horizontal line corresponds to the
range of $r$ values for a bound terminating orbit.}\label{potential_polynomial7}
\end{center}
\end{figure}

This case corresponds to the choice $V_{\rm eff}^{\rm min} = \frac{E^2-\varepsilon}{2}$.
The effective potential and the corresponding characteristic polynomial $P(z)$ are shown
in Fig. \ref{potential_polynomial7}. The line $\frac{E^2-\varepsilon}{2}$ intersects the effective potential
$V_{\rm eff}(r)$ at $r=6m$. 
This corresponds to $e_1=e_2=e_3=0$. We are allowed to integrate from $e_1=e_2=e_3$ to $z=\infty$, which corresponds
to integration from $r_1=r_2=r_3=6m$ to $r=\infty$.

\subsubsection{Bound terminating orbit}
\begin{figure}[htbp]
\begin{center}
\resizebox{6in}{!}{\includegraphics{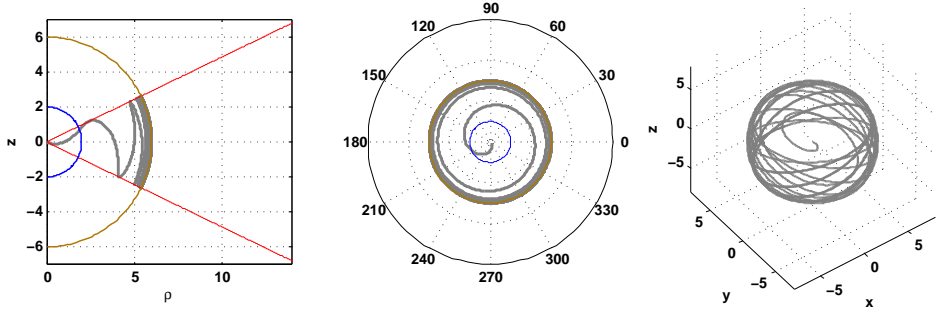}}
\caption{We show a bound terminating orbit for a massive test particle ($\varepsilon=1$) in the $\rho$-$z$-plane
(with $\rho=r\sin\theta$, $z=r\cos\theta$) (left), in the plane perpendicular to $\vec{L}$ (middle) and in $\mathbb{R}^3$ (right).
We have chosen $\beta=0.9$, $E=\sqrt{8/9}$, $L_{z}=\sqrt{12\times (0.9)^2}$, $k=\frac{10}{9}$ and $m=1$. The dark yellow and blue circles in the $\rho$-$z$-plane 
and in the plane perpendicular to $\vec{L}$ correspond to circles with radius $r=6m=6$ and the Schwarzschild radius $r_s=2m=2$, respectively.}\label{mszrdiscb090EminLmin}
\end{center}
\end{figure}

The range of allowed $r$ values is indicated by the horizontal red line in Fig.\ref{potential_polynomial7}
and is  $0 \le r\le 6m$.
An example of a bound terminating orbit is given in Fig.\ref{mszrdiscb090EminLmin} for $\beta=0.9$.

\subsubsection{Spherical orbit}
\begin{figure}[htbp]
\begin{center}
\resizebox{6in}{!}{\includegraphics{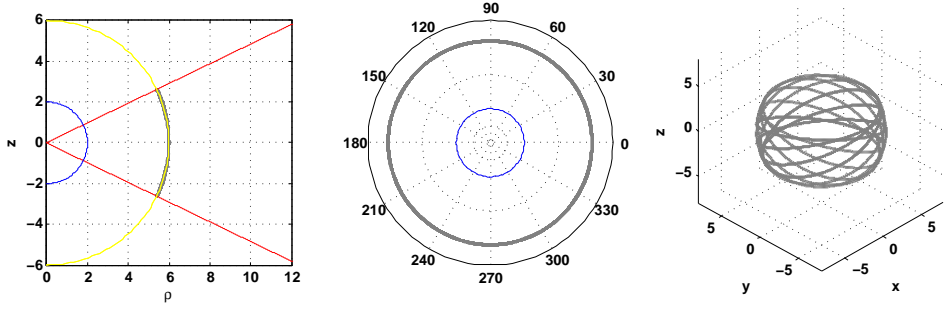}}
\caption{We show a spherical orbit for a massive test particle ($\varepsilon=1$) in the $\rho$-$z$-plane
(with $\rho=r\sin\theta$, $z=r\cos\theta$) (left), in the plane perpendicular to $\vec{L}$ (middle) and in $\mathbb{R}^3$ (right).
We have chosen $\beta=0.9$, $E=\sqrt{8/9}$, $L_{z}=\sqrt{12\times (0.9)^2}$, $k=\frac{10}{9}$ and $m=1$. The dark yellow and blue circles in the $\rho$-$z$-plane 
and in the plane perpendicular to $\vec{L}$ correspond to circles with radius $r=6m=6$ and the Schwarzschild radius $r_s=2m=2$, respectively. }\label{mszrdiscb090EminLmin2c}
\end{center}
\end{figure}

For $r_0= 6m$ we have a spherical orbit. These spherical orbits have already been discussed
in \cite{gm}. An example of a spherical orbit is given in Fig.\ref{mszrdiscb090EminLmin2c} for $\beta=0.9$.

\subsection{Geodesics for $D=0$ and $E^2-\frac{2}{3}\varepsilon < \frac{L^2}{54m^2}$}
\begin{figure}[htbp]
\begin{center}
\resizebox{6in}{!}{\includegraphics{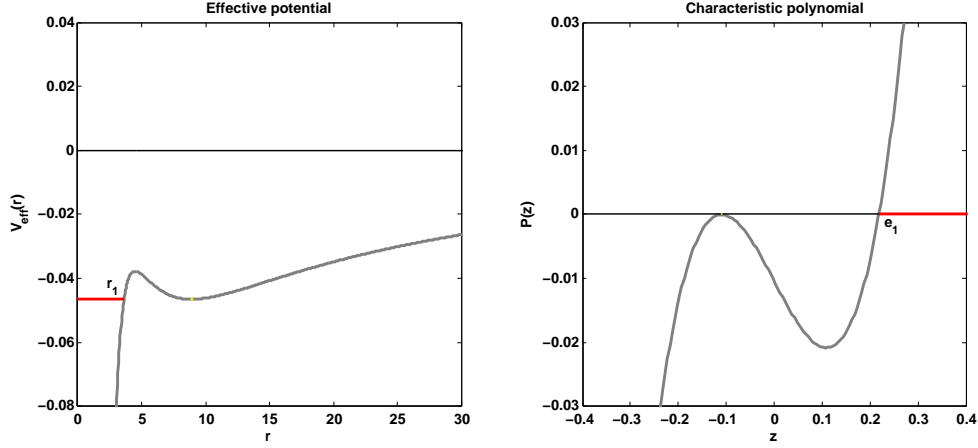}}
\end{center}
\caption{The effective potential $V_{\rm eff}(r)$ (left) and the characteristic polynomial $P(z)$ (right) for $D=0$
and $E^2-\frac{2}{3}\varepsilon< \frac{L^2}{54m^2}$. The red horizontal line corresponds to the
range of $r$ values for a bound terminating orbit.}\label{potential_polynomial9}
\end{figure}

The effective potential and the corresponding characteristic polynomial $P(z)$ are shown
in Fig. \ref{potential_polynomial9}. The line $\frac{E^2-\varepsilon}{2}$ intersects the effective potential
$V_{\rm eff}(r)$ at $r_1$ and at the minimum of the effective potential at $r=r_{2}$.
 These intersection points corresponds to a simple zero $e_1$ and a double zero at $e_{2}=e_3$.
We are only allowed to 
integrate between $\infty$ and $e_1$, which corresponds to integration from $r=0$ to $r=r_1$.

\subsubsection{Bound terminating orbit}

\begin{figure}[htbp]
\begin{center}
\resizebox{6in}{!}{\includegraphics{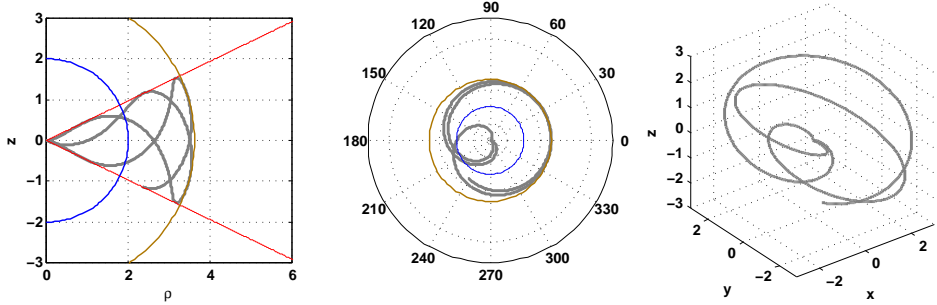}}
\end{center}
\caption{We show a bound terminating orbit for a massive test particle ($\varepsilon=1$) in the $\rho$-$z$-plane
(with $\rho=r\sin\theta$, $z=r\cos\theta$) (left), in the plane perpendicular to $\vec{L}$ (middle) and in $\mathbb{R}^3$ (right).
We have chosen $\beta=0.9$, $E=0.95229$, $L_{z}=3.3$ and $k=\frac{10}{9}$.}\label{mszrdiscb090E09523Lz33BTeps}
\end{figure}

An example of a bound terminating orbit is shown in Fig.\ref{mszrdiscb090E09523Lz33BTeps}.

\subsubsection{Spherical orbit}
Choosing $r_0=r_{2}$ we find a  spherical orbit with radius $r_{2}$. In this case, the spherical
orbit is a stable spherical orbit. The qualitative plot is similar to the plot given in Fig.\ref{mszrdiscb090EminLmin2c}.

\section{Newtonian limit}
Assuming the gravitational field to be weak and the test particles to move slowly, we can
give the Newtonian limit of our equations. This reads \cite{carroll}:
\begin{equation}
 \frac{d\vec{r}^2}{dt^2}=\frac{1}{2} \vec{\nabla} h_{00}
\end{equation}
where we assume $g_{\mu\nu}=\eta_{\mu\nu}+h_{\mu\nu}$ with $h_{\mu\nu}$ very small
and $\eta_{\mu\nu}$ to be the Minkowski metric.  Proceeding as in classical mechanics, we find the Virial theorem
\begin{equation}
 -2 < T > = \frac{1}{2} < \vec{r}\cdot \vec{\nabla}h_{00}> =
\frac{1}{2} < r \partial_r h_{00} >
\end{equation}
where $< .. >$ denotes the temporal average and $T=\frac{1}{2}\left(\frac{d\vec{r}}{dt}\right)^2$ is the kinetic energy. Hence, the Virial
theorem doesn't change for $\beta\neq 1$ as compared to the Newtonian limit of the standard Schwarzschild case $\beta=1$. However, note that the kinetic energy $T$ has a $\beta$-dependence 
since
\begin{equation}
 \left(\frac{d\vec{r}}{dt}\right)^2= \left(\frac{dr}{dt}\right)^2 + r^2 \left(\frac{d
\theta}{dt}\right)^2 + r^2 \sin^2\theta \beta^2 \left(\frac{d\phi}{dt}\right)^2 \ .
\end{equation}
Since the average of the potential energy $< r \partial_r h_{00} >$ does not depend on $\beta$ and the Virial theorem
holds for all $\beta$ this means that when $\beta < 1$ the angular
velocity $\left(\frac{d\phi}{dt}\right)^2$ increases. We have -- in fact -- seen this
effect 
when studying the perihelion shift of planets. For the same
interval of eigentime, the planet moves
further in a space-time with $\beta < 1$ as compared to in a space-time with $\beta=1$. 

In addition, it is easy to show that Kepler's third law $T^2/a^3=const.$, where $T$ is the period of the orbit and $a$ the
semimajor axis also doesn't change. This is easiest to see when considering
circular and planar orbits with $L_z=L$, i.e. $\theta=\pi/2$.
For these, we let $\dot{r}=0$ and substitute $L_z/r$ from $\dot{\phi}$ into
$\dot{r}$ (see (\ref{rcompo}) and (\ref{thetacompo}) ). After separation
the two sides can be integrated to give the period of the orbit.
The relation will contain
a dependence on $\beta$. However, note that the integration in $\phi$ is from
$0$ to $2\pi\beta$ such that the $\beta$ cancels and $T$ and $a$ are independent of $\beta$.

\section{Conclusions and Outlook}
In this paper, we have constructed the complete set of geodesics in the space-time
of a Schwarzschild black hole pierced by a cosmic string.
We have classified the geodesics according to their energy, angular momentum aligned with the axis of the string $L_z$, the ratio between total angular momentum and $L_z$, the horizon
radius of the black hole and the deficit angle (or energy per unit length of the string).
We found that the change of the deficit angle doesn't effect the $r(\theta)$ motion, but that the $r(\phi)$ motion changes significantly. Moreover, the motion
of the particles is in general not planar.
We found a bound of the energy per unit length of the string comparing our results
with experimental tests of General Relativity. From perihelion shift and light deflection
we found $\mu \lesssim 10^{17} \frac{\rm kg}{\rm m}$ and $\mu \lesssim 10^{16} \frac{\rm kg}{\rm m}$, respectively.
The existence of cosmic strings may also influence the creation of gravitational waves since for inspirals and
flybys the deficit angle modifies the angular velocity and, thus, the temporal change of the effective quadrupole responsible for the emitted gravitational wave.
The results obtained in this paper can be generalized to rotating black holes, i.e.
to the space-time of a Kerr black hole pierced by a cosmic string. They can further be generalized to
a Schwarzschild--de Sitter black hole pierced by a cosmic string. For treating such orbits one has to
apply hyperelliptic integration developed in \cite{hl1}. We expect the same enlarged variety
of orbits as we encountered in Schwarzschild--de Sitter space--time as compared to Schwarzschild space--time.
\\
\\
{\bf Acknowledgements}
We thank the DFG and Jacobs University Bremen for financial support. We also thank A. Aliev for bringing
reference \cite{ag} to our attention.


\begin{thebibliography}{34}
\bibitem{hagihara} Y. Hagihara, {\it Theory of relativistic trajectories in a gravitational field of Schwarzschild}, \emph{Japan. J. Astron. Geophys}. \textbf{8}, 67 (1931).
\bibitem{chandra} S.Chandrasekhar, {\it The Mathematical Theory of black holes},
Oxford University Press (1998).
\bibitem{hl1} E.~Hackmann and C. L\"ammerzahl, 
 Phys. Rev. D {\bf 78} 024035 (2008); Phys. Rev. Lett.  {\bf 100}, 171101 (2008).
\bibitem{hackmann08} E.~Hackmann, V.~Kagramanova, J.~Kunz and C.~L\"ammerzahl,
  Phys.\ Rev.\  D {\bf 78} 124018 (2008);
  [Erratum-ibid.\  {\bf 79} (2009) 029901]
\bibitem{hackmannetal1} E.~Hackmann, V. ~Kagramanova, J.~Kunz and C.~L\"ammerzahl, {\it Analytic
solutions of the geodesic equation in Kerr--(anti-)de Sitter space--times}, Phys. Rev. D, to appear; E.~Hackmann, C.~L\"ammerzahl, and A.~Macias, {\it Complete classification of 
geodesic motion in fast Kerr and Kerr--(anti-)de Sitter space--times}, In:{\it New trends
in statistical physics: Festschrift in honour of Leopoldo Garcia-Colin's 80th birthday} (World Scientific, Singapore 2010), to appear.
\bibitem{hackmannetal2} E.~Hackmann, V. ~Kagramanova, J.~Kunz and C.~L\"ammerzahl, Europhys. Lett. {\bf 88}, 30008 (2009).
\bibitem{polchinski} see e.g. J. Polchinski, {\it Introduction to cosmic
F- and D-strings}, hep-th/0412244 and reference therein.
\bibitem{vs} A. Vilenkin and  P. Shellard, {\it Cosmic strings and other topological defects}, Cambridge University Press (1994).
\bibitem{braneinflation} M. Majumdar and A. Davis, JHEP {\bf 03}, 056 (2002); S. Sarangi and S. Tye, Phys. Lett.B {\bf 536}, 185 (2002). 
\bibitem{bw} R. Bach and H. Weyl, Math. Zeit. {\bf 13}, 134 (1922).
\bibitem{afv} M. Aryal, L. Ford and A. Vilenkin, Phys. Rev. D {\bf 34}, 2263 (1986)
\bibitem{dgt} F. Dowker, R. Gregory and J. Traschen, Phys. Rev. D {\bf 45}, 2762 (1992).
\bibitem{agk} A.~Achucarro, R.~Gregory and K.~Kuijken,
  Phys.\ Rev.\  D {\bf 52}, 5729 (1995).
\bibitem{fbl} 
  W.~Freire, V.~Bezerra and J.~Lima,
  Gen.\ Rel.\ Grav.\  {\bf 33} 1407 (2001).
\bibitem{ag} A. Aliev and  D. Gal'tsov, Piz'ma Astron. Zh. {\bf 14},
116 (1988); {\it translated in} Sov. Astron. Lett. {\bf 14},  48 (1988).
\bibitem{gm} D. Gal'tsov and E. Masar, Class. Quantum Grav. {\bf 6}, 1313 (1989).
\bibitem{cb} S. Chakraborty and L. Biswas, Class. Quantum Grav. {\bf 13}, 2153 (1996).
\bibitem{Ab}  M. Abramowitz, I. A. Stegun, 1972. \ Handbook of Mathematical Functions with Formulas, Graphs, and Mathematical Tables, 9th printing. New York: Dover, pp. 627-671.
\bibitem{valtonen} M. J. Valtonen et. al, Nature {\bf 452}, 851 (2008).
\bibitem{will1} {\it see e.g.} C. Will, {\it Theory and Experiment in Gravitational physics}, Cambridge University Press (1993).
\bibitem{turyshev} {\it for a summary of recent results see e.g.} S. Turyshev, Ann. Rev. Nucl. Part.Sci.{\bf 58}, 207 (2008)
\bibitem{wtb} J. Williams, S. Turyshev and D. Boggs, Phys. Rev. Lett. {\bf 93}, 261101 (2004).
\bibitem{bit} B. Bertotti, L. Iess and P. Tortora, Nature {\bf 425}, 374 (2003). 
\bibitem{loc} J. Levin, R. O'Reilly and E. Copeland, (Sussex U.), Phys. Rev. D {\bf 62} 024023 (2000). 
\bibitem{GmBzMl} M.G. Germano, V.B. Bezerra, E.R. Bezerra de Mello (1997), Class. Quantum Grav \textbf{13} 2663 (1997)
\bibitem{carroll} {\it see e.g.} S. Carroll, {\it Spacetime and Geometry: An Introduction to General Relativity}, Benjamin Cummings (2003).
\end{thebibliography}
\end{document}